\documentclass[]{aa}

\usepackage{multirow}
\usepackage{multicol}

\usepackage{graphicx}
\usepackage{txfonts}

\usepackage{color}
\usepackage[colorlinks=true,allcolors=blue]{hyperref}
\usepackage{natbib}
\bibpunct{(}{)}{;}{a}{}{,} 
\usepackage{paralist}
\usepackage{epstopdf}
\usepackage{epsfig}
\usepackage{amsmath}
\usepackage{tablefootnote}
\usepackage{amssymb}
\usepackage{bm}
\usepackage{tabularx}
\usepackage{lmodern}

\begin{document} 

    \title{Exploring multiple stellar populations in globular clusters with Euclid: a theoretical overview and insights from NGC 6397} 
       \author{A.\,P.\,Milone$^{1,2}$\thanks{E-mail: antonino.milone@.unipd.it},
        G.\,Cordoni$^{3}$,
        A.\,F.\,Marino$^{2}$,
        V.\,Altomonte$^{1}$,
        E.\,Dondoglio$^{2}$,
        M.\,V.\,Legnardi$^{1}$,
        E.\,Bortolan$^{1}$, 
        S.\,Lionetto$^{1}$,
        A.\,V.\,Marchuk$^{1}$,
        F.\,Muratore$^{1}$,
        T.\,Ziliotto$^{1}$
        }
\institute{$^1$  Dipartimento di Fisica e Astronomia ``Galileo Galilei'', Univ. di Padova, Vicolo dell'Osservatorio 3, Padova, 35122, Italy \\
$^{2}$ Istituto Nazionale di Astrofisica - Osservatorio Astronomico di Padova, Vicolo dell'Osservatorio 5, Padova, 35122, Italy\\
$^{3}$ Research School of Astronomy \& Astrophysics, Australian National University, Canberra, ACT 2611, Australia \\
}

\titlerunning{Multiple populations in globular clusters with Euclid} 
\authorrunning{Milone et al.}

\abstract{We investigate the behavior of multiple stellar populations in globular clusters (GCs) using photometric diagrams constructed with Euclid photometry. By employing synthetic spectra and isochrones that incorporate the chemical differences between first-population (1P) stars, resembling field stars, and second-population (2P) stars, enriched in helium and nitrogen but depleted in carbon and oxygen, we identify, from a theoretical perspective, the color-magnitude diagrams and the chromosome maps most effective at distinguishing these populations within GCs. Euclid photometry proves to be a powerful tool for identifying multiple populations among M-dwarfs, as 1P and 2P stars form distinct sequences in well-chosen photometric diagrams, driven by differences in the strength of oxygen-based molecular features, such as water vapor. To validate our theoretical findings, we analyzed Euclid photometry and astrometry of the GC NGC\,6397, complemented by photometric and astrometric data from the {\it Hubble Space Telescope} and {\it James Webb Space Telescope}, enabling a comprehensive study of its stellar populations across a wide field of view. 
We find that the 1P constitutes $\sim$30\% of the M-dwarfs in NGC\,6397, with the fraction of 1P stars remaining consistent across different stellar masses and throughout the entire field of view. 2P stars exhibit an [O/Fe] depletion of about 0.3 dex relative to 1P stars, and both populations display isotropic proper motions. This study represents the first comprehensive analysis of multiple populations among M-dwarfs across a wide field of view, demonstrating that Euclid photometry is a powerful instrument for investigating multiple populations in GCs.
}
 
\keywords{  globular clusters: general, stars: population II, stars: abundances, techniques: photometry.}

\maketitle

\section {Introduction}
\label{sec:intro}

Despite being long-held as examples of simple stellar populations, extensive research carried out in the past decades has shown that the majority Galactic globular clusters (GCs) are actually host to distinct stellar populations with significant chemical variations. These can be grouped into a first population (1P), whose chemical makeup resembles field stars of similar metallicity, and one or more second populations (2P), which feature higher abundances of helium, nitrogen, aluminum and sodium, and decreased carbon and oxygen \citep[see, e.g.\,][for reviews]{kraft1994a, bastian2018a, gratton2019a, milone2022a}.

The existence of multiple, discrete populations in GCs, their characteristics and their variation between clusters, represents a significant challenge to our models of stellar formation and evolution in the early Universe, and the contrasting scenarios proposed to explain them paint very different pictures not just of cluster origin, but potentially of Galactic-halo assembly and even Universe reionization \citep[e.g.][]{renzini2017a}. For this reason, they have been the focus of a large-scale observational effort over several decades.

The search for multiple stellar populations in GCs from photometry has primarily focused on UV filters, which are sensitive to spectral features influenced by molecular bands such as carbon, nitrogen, and oxygen \citep[e.g.][]{marino2008a, yong2008a, milone2012a, lee2022a, mehta2025a}. The densely populated central regions of clusters are typically studied using the high-resolution capabilities of the {\it Hubble Space Telescope} ({\it HST}), while the more extended outskirts are explored with wide-field ground-based facilities
\citep[][for early studies on multiple populations in GCs based on the sinergy of {\it HST} and ground-based telescopes]{milone2012a, milone2013a}. 

While effective, this approach is constrained by the limitations of UV detectors in achieving high-precision photometry for faint stars. Consequently, these studies typically focus on giant stars or bright main-sequence (MS) stars. 
Optical and near-infrared filters are poorly sensitive to star-to-star carbon, nitrogen, and oxygen variations contents among bright stars. 
However, they are effective tools for identifying stellar populations with different oxygen abundances among M-dwarfs. This was demonstrated by \citet{milone2012b} using photometry of the GC NGC\,2808 in the F110W and F160W bands of the infrared channel of the Wide Field Camera 3 (IR/WFC3) on board {\it HST}.
The 2P stars, which have lower oxygen content than 1P stars, exhibit spectra that are less absorbed by oxygen-containing molecules. Since the F160W band is particularly sensitive to absorption by these molecules, 2P stars display brighter F160W magnitudes and bluer F110W$-$F160W colors compared to 1P stars of similar luminosity \citep[see also][for studies on multiple populations among low-mass stars based on {\it HST} and {\it James Webb} space telescope, {\it JWST} ]{milone2014a, milone2017a, milone2019a, milone2023b, dondoglio2022a, ziliotto2023a, cadelano2023a, scalco2024a, scalco2024b, marino2024a, marino2024b}.

A significant limitation of studies conducted with photometric cameras aboard the {\it HST} and {\it JWST} is their relatively small field of view. In contrast, the recently launched Euclid Telescope by the European Space Agency (ESA) addresses this challenge with its considerably wider field of view, enabling the study of larger areas of the sky in a single observation. 

In this paper, we explore the potential of Euclid photometry to detect and characterize multiple stellar populations in GCs. Our analysis begins with the use of isochrones tailored to the specific chemical compositions of 1P and 2P stars in Euclid filters. We then examine Euclid photometry of the GC NGC\,6397, analyze its stellar populations, and compare the resulting photometric diagrams with those derived from {\it HST} and {\it JWST} observations.

NGC\,6397, the second-nearest GC, has been extensively studied in the context of multiple stellar populations. Observations have revealed two distinct stellar sequences along both the red-giant branch (RGB) and the MS, as identified in CMDs constructed using photometry from {\it HST} and {\it JWST} \citep{milone2012c, milone2017b, scalco2024a}. These sequences correspond to chemically distinct populations, exhibiting moderate differences in helium, oxygen, and sodium abundances \citep[][]{lind2011a, carretta2009a, milone2018a, marino2019a}. The small elemental abundance differences between 1P and 2P stars, combined with the low metallicity of NGC\,6397 ([Fe/H] $\sim -2.02$, based on the 2010 edition of the \citet{harris1996a} catalog), result in only minor flux variations among stars with similar atmospheric parameters. This often makes it challenging to photometrically distinguish between 1P and 2P stars \citep[e.g.,][]{nardiello2015a, branco2024a, mehta2025a}.

The paper is organized as follows: 
    Section\,\ref{sec:data} describes the dataset and summarizes the methods used for data reduction.
    Section\,\ref{sec:theory} focuses on the theoretical modeling of multiple populations.
    Section\,\ref{sec:ngc6397} presents the observational results for NGC\,6397.
    Finally, Section\,\ref{sec:conclusions} provides a summary of our findings and conclusions.

\section{Data and data reduction}
\label{sec:data}
\begin{figure*} 
  \centering
 \includegraphics[height=10cm,trim={0.0cm 0.0cm 0.0cm 0.0cm},clip]{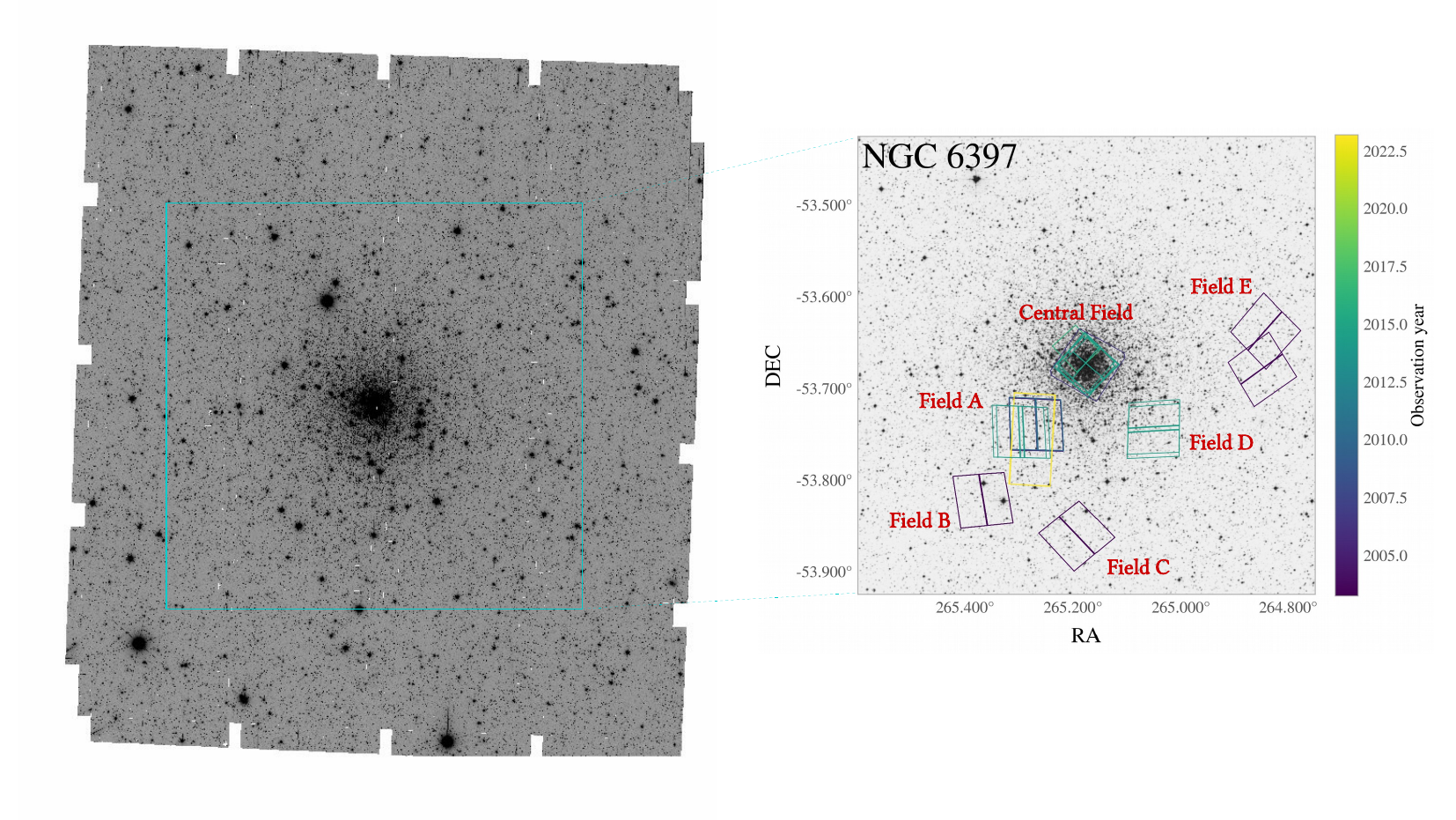}
 \includegraphics[height=8cm,trim={0.0cm 0.0cm 0.0cm 0.0cm},clip]{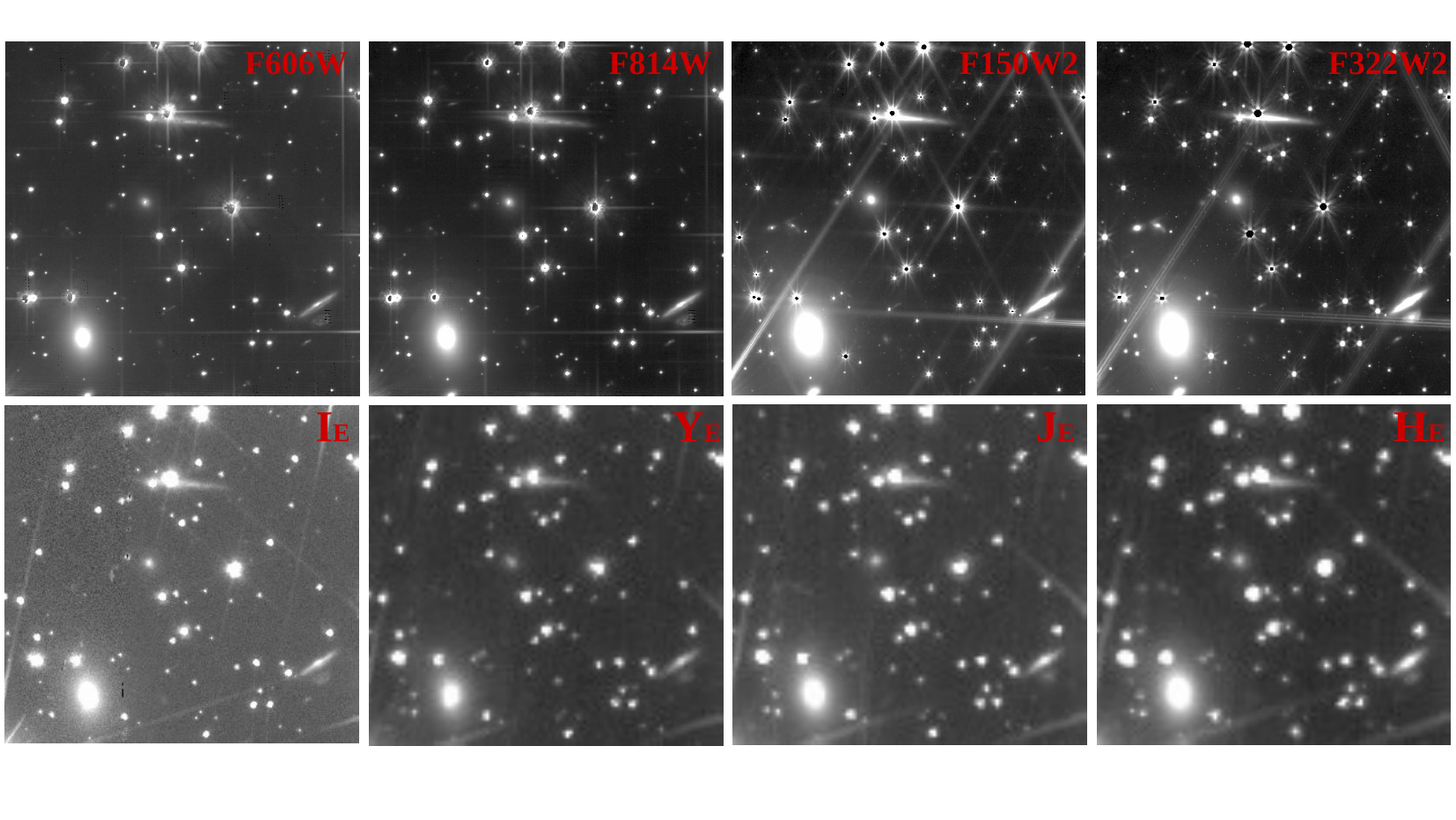}
  \caption{Stacked image of the Euclid images in the H$_{E}$ band \citep{libralato2024a}. The right panel shows a zoom of the studied field of view together with the footprints of the {\it HST} and {\it JWST} images used in this paper. The color scale is associated with the epoch as indicated by the color-bar on the right. The bottom panels compare the observations of a 30 square arcsec region obtained in the F606W and F814W bands of WFC/ACS, the F150W2 and F322W2 filters of NIRCam and the Euclid filters.}  
  \label{fig:footprint}
\end{figure*}

To evaluate the potential of Euclid for identifying and characterizing multiple stellar populations in GCs, we analyzed photometry and astrometry of stars within the field of view of NGC\,6397. This analysis was complemented with additional data from Gaia DR3 \citep{gaia2021a}, {\it JWST}, and {\it HST}.
 The footprints of the {\it JWST} and {\it HST} images, shown in  Figure\,\ref{fig:footprint}, include six fields: the central field and fields A to E. Most observations are concentrated in the central field and field A.

The central field includes images from WFC/ACS \citep[GO-10755,][]{sarajedini2007a, anderson2008a} and UVIS/WFC3 \citep[GO-13297,][]{piotto2015a}, used to identify 1P and 2P stars along the MS and RGB. Infrared images from WFC3 \citep[IR/WFC3,][]{correnti2018a} analyze multiple populations among M-dwarfs, supplemented by UVIS/WFC3 and WFC/ACS data for proper motions.

Field A in Figure\,\ref{fig:footprint} is located about 5 arcmin south-east the cluster center and has been extensively observed with ACS/WFC of {\it HST} \citep[GO-10424 and GO-11633][]{richer2006a} and NIRCam of {\it JWST} \citep{bedin2024a} to study faint stars in NGC\,6397. 
The remaining fields B, C, D, and E are observed with ACS/WFC only.
The key properties of the images utilized in this study are summarized in Table\,\ref{tab:data} and are discussed in the following.
As an example of the quality of the images collected using different telescopes and filters, the bottom panels of Figure\,\ref{fig:footprint} compare the stacked images obtained with {\it HST}, {\it JWST}, and Euclid for a 30-square-arcsecond region within field A.

\subsection{Euclid}
The primary dataset comprises stellar magnitudes and positions obtained by \citet{libralato2024a} from Euclid , a space observatory launched in 2023 and designed for surveys. Light is collected by a 1.2m Korsh telescope, and directed by a dichroic mirror to two onboard instruments: The VISible Instrument (VIS) and the Near-Infrared Spectrometer and Photometer (NISP).
VIS consists of a square array of 36 CCDs, each equipped with a matrix of 4K $\times$ 4K 12 $\mu$m pixels, for a total field of view of 0.54 deg$^2 $ and a scale of 100 mas\,pixel$^{-1} $. As the name suggests, it is meant for photometry in the visible range and features a single broadband filter ($I_{E}$) with a wavelength range from 550 to 900 nm. NISP, meanwhile, features both a spectroscopic channel (not used in this work) and a photometric one. It is composed by a 4$\times$4 array of TIS detectors featuring $2K \times 2K$ $18 \mu m$ pixels each, for a total FoV of 0.57 deg$^2 $ and a pixel scale of 300 mas\,pixel$^{-1}$. Three NIR filters ($Y_E$, $J_E$ and $H_E$) cover the wavelength ranges of 920-1146 nm, 1146-1372 nm and 1372-2000 nm, respectively, allowing for multiband photometry \citep{schirmer2022a}. A detailed description of VIS and NISP can be found in the Euclid collaboration papers by \citet{euclid_cropper2024a} and \citet{euclid_jahnke2024a}. 
As put forward by \cite{massari2024a} and \cite{libralato2024a} Euclid's high resolving power (comparable with that of {\it HST} in their overlapping wavelength range) and wide FoV give it a high potential for the study of extended, crowded objects such as GCs.

The Euclid NGC\,6397 dataset used in this work is based on the Reference Observation Sequence
 \citep[ROS,][]{euclid_scaramella2022a} carried out by Euclid on September 22nd, 2023 as part of the Euclid Early Release Observations (ERO) program. The ROS, centered on the cluster, consists in four dithered observations, each including three 87.2s exposures in the $Y_E$, $J_E$ and $H_E$ filters, a spectral exposure (not featured in this study) and a 560s long exposure in the $I_E$ broadband filter. Each individual, non-stacked exposure was reduced by \cite{libralato2024a}. As a first step, the raw images were corrected according to the standard ERO pipeline detailed in \cite{cuillandre2024a} but without correcting VIS data for the effects of cosmic rays. Then, each image was reduced by means of effective-PSF photometry using \texttt{euclid1pass}, a version of the \texttt{hst1pass} software developed by \cite{anderson2022a} adapted to Euclid specifications. The results were then corrected for geometric distortion following the procedures for wide-field images described in \cite{libralato2015a} using star positions from Gaia DR3 as a reference, calibrated and aligned to the same reference frame. This yielded increased astrometric and photometric precision when compared to the official ERO catalog.

\subsection{HST and JWST}\label{subsec:hst}
The photometry and astrometry of stars in the {\it HST} and {\it JWST} fields were performed using computer programs designed and developed by Jay Anderson \citep[e.g.,][]{anderson2000a, anderson2008a, anderson2022a}. Specifically, we utilized the KS2 program, an evolution of the kitchen\_sync program \citep{anderson2008b}, which was originally developed to extract high-precision stellar fluxes and positions from images obtained with the Wide-Field Channel of the Advanced Camera for Surveys (WFC/ACS) aboard {\it HST}.

KS2 is a versatile program for deriving precise stellar magnitudes and positions, utilizing three methods tailored to stars with varying brightness and in diverse stellar environments: \begin{itemize}
    \item     Method I targets bright stars, identifying those with prominent peaks within a 5$\times$5-pixel area after subtracting nearby sources. Fluxes and positions are calculated in each image using a PSF model specific to the star's location, while the sky level is determined from the surrounding region (4–8 pixels from the center). Measurements across all exposures are averaged to ensure high precision.

    \item Method II is designed for fainter stars, performing weighted aperture photometry within a 5$\times$5 grid. Pixels influenced by neighboring stars are down-weighted, and the sky level is measured similarly to Method I.

    \item Method III suits faint stars in crowded fields with many exposures. Photometry is performed within a smaller 0.75-pixel-radius aperture, and the sky is calculated closer to the star (2–4 pixel annulus).
\end{itemize}

All methods involve independent measurements from individual exposures, which are averaged for the most accurate results. KS2 also includes diagnostics to ensure only isolated stars with high-quality fits are used for final analyses \citep[see][for details on the selection procedures]{milone2023a}.

Stellar coordinates were corrected for geometric distortion using the solutions from \citet{anderson2006a, bellini2009a, bellini2011a, anderson2022a} and \citet{milone2023b}. Photometric calibration to the Vega magnitude system followed \cite{milone2023a}, utilizing the latest zero points available on the STScI website.
Moreover, we used the KS2 program artificial-star tests to evaluate the photometric uncertainties on stellar magnitudes \citep[see][for details]{anderson2008a, milone2023a}.

\subsection{Proper motions}
To measure relative proper motions, we followed a procedure outlined in several studies \citep[e.g.,][]{anderson2003a, piotto2012a, libralato2022a, milone2023a}, which relies on comparing stellar positions across images taken at different epochs.

Initially, we identified distinct photometric and astrometric catalogs created from images captured in various filters over multiple epochs. The reference frame for our analysis was anchored to the first-epoch images taken in the reddest filter, which served as the master frame. We adopted a specific orientation for the reference frame, with the X-axis aligned toward the west and the Y-axis pointing north.

Star coordinates from each catalog were transformed into this master frame using six-parameter linear transformations. To further refine the alignment and reduce the impact of small residual distortions, we applied local transformations based on the nearest 75 reference stars. The target stars themselves were excluded from contributing to the transformations used to correct their positions.

The abscissa and ordinate of each star were plotted as functions of the observation epoch and fitted using a weighted least-squares straight line. The slope and the corresponding uncertainty of this line provide the best estimate of the star's proper motion.

The transformations rely on bright, unsaturated cluster stars that meet the selection criteria outlined in Section,\ref{subsec:hst} and \cite{milone2023a}. These stars were selected through a two-step process. First, we identified probable members of NGC\,6397 based solely on their position in the CMD and used them to derive preliminary proper motions. In the second step, we refined the selection by considering each star's location in both the CMD and the proper motion diagram, allowing for a more accurate determination of cluster membership. These refined members were then used to compute improved proper motions. To minimize uncertainties in proper-motion measurements, we corrected the stellar coordinates at each epoch by subtracting the displacements caused by proper motions. These adjusted coordinates were then used to compute improved transformations, further refining the proper-motion estimates.

The limited availability of archival {\it HST} and {\it JWST} first-epoch images restricts the possibility to disentangle field stars from cluster members to a relatively small area. To expand the field of view, we utilized the photometric catalog of NGC\,6397 compiled by Peter Stetson \citep{stetson2000a, stetson2019a}, previously employed in studies of multiple populations in this cluster by \citet{monelli2013a}.

Specifically, we computed the displacements, DX and DY, of stellar coordinates in Stetson's and Libralato's catalogs relative to NGC\,6397 stars, using six-parameter linear transformations. It is important to note that the stellar astrometry in the catalogs by \citet{monelli2013a} and \citet{stetson2019a} is based on images collected from various telescopes over different epochs and has not been rigorously corrected for geometric distortion. As a result, the derived stellar displacements are unsuitable for investigating the internal kinematics of cluster stars.

However, the significant proper motion of NGC\,6397 relative to field stars enables a clear distinction between the two populations based on stellar displacements. This is demonstrated in Figure\,\ref{fig:ngc6397_rmap}, where the CMD of stars from Euclid photometry is shown (middle panel), along with $I_{\rm E}$ plotted against the displacement $\delta$R=$\sqrt{DX^{2}+DY^{2}}$ (right panel). We classified stars with $\delta$R$<$1.5 as probable cluster members, while those with larger displacements were identified as field stars. In the middle and right panels of Figure\,\ref{fig:ngc6397_rmap}, cluster members are shown in black, and field stars are depicted in gray.

\subsection{Differential reddening}
The photometry has been corrected for differential reddening using a procedure similar to that introduced by \citet{milone2012a}, which has been widely adopted in various studies on reddening by our team \citep{jang2022a, milone2023a, legnardi2023a}.

To analyze differential reddening from Euclid photometry, we constructed $H_{E}$ versus X$-H_{E}$ diagrams, where X = $I_{E}$, $Y_{E}$, and $J_{E}$. Each of these diagrams was utilized to extract reddening information from a carefully chosen set of reference cluster members. These reference stars were selected from regions of the CMD where the reddening vector forms a significant angle with the cluster's fiducial line. This selection ensures that the effects of differential reddening on stellar colors and magnitudes can be distinguished clearly from shifts caused by photometric uncertainties.

We transformed each CMD into a new reference frame by rotating the original diagram such that the axes align with the reddening vector. In this new frame, the abscissa runs parallel to the reddening direction, while the ordinate is orthogonal to it. The adopted absorption coefficients for the Euclid filters are listed in Table\,\ref{tab:coefficienti} and obtained as in \cite{legnardi2023a}, by assuming a standard reddening law with R$_{V}$=3.1 and E(B$-$V)=0.18 mag \citep[2010 version of the][catalog]{harris1996a}.

\begin{table}
  \caption{Relative extinction, A/E(B$-$V), for Euclid bandpasses.}
\begin{tabular}{c c c c }
\hline
 $I_{\rm E}$ & $Y_{\rm E}$ & $J_{\rm E}$  & $H_{\rm E}$  \\
 2.498       & 1.223       &   0.867      &  0.528        \\
     \hline
\end{tabular}
  \label{tab:coefficienti}
 \end{table}

We determined the abscissa difference, $\Delta(X-H_E)$, between each reference star and the main-sequence (MS) fiducial line in each rotated CMD. The measured $\Delta(X-H_E)$ values were compared to those predicted for reddening variations ranging from $\Delta E(B-V) = -0.1$ to $0.1$ mag, sampled in increments of 0.001 mag. The reddening variation, $\Delta E(B-V)$, that minimizes the $\chi^2$ value was taken as the optimal differential-reddening estimate for the corresponding reference star.
The most reliable differential reddening value was derived as the median of the $\Delta E(B-V)$ values of the 75 spatially nearby reference stars. 
We adopted the same procedure discussed above to correct the effects of differential reddening from ACS/WFC, UVIS/WFC3 and NIRCam photometry. 

\subsection{Results}
\begin{figure*} 
  \centering
 \includegraphics[height=7.75cm,trim={0.0cm 0.0cm 0.0cm 0.0cm},clip]{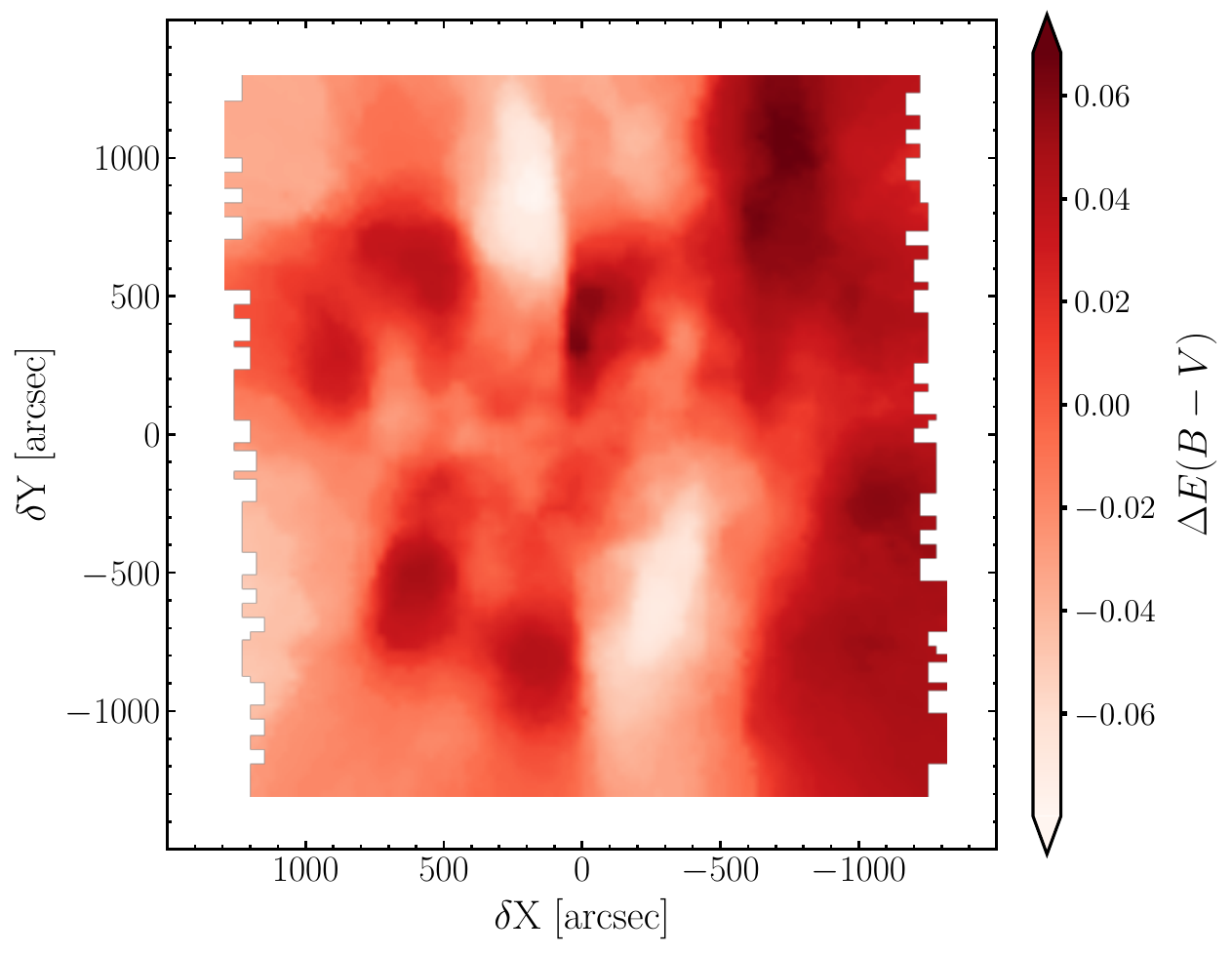}
 \includegraphics[height=9.5cm,trim={0.0cm 5.5cm 4.0cm 0.0cm},clip]{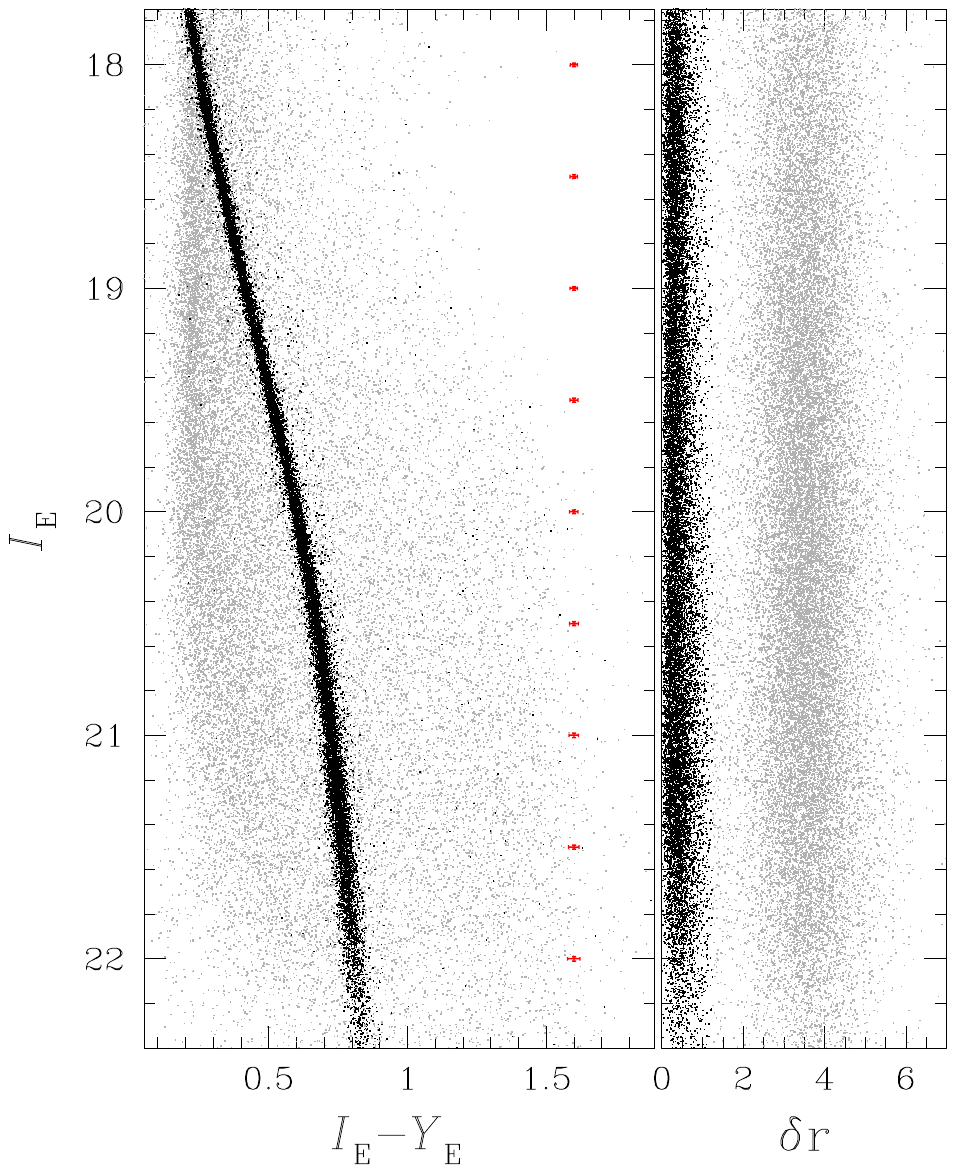}
  \caption{Map of differential reddening toward NGC\,6397.
  The color levels in the map correspond to the relative 
  variation in E(B$-$V), with the scale indicated on the right (left panel). 
  The middle panel presents the $I_{\rm E}$ vs.\,$I_{\rm E}-Y_{\rm E}$ CMD corrected for differential reddening. The right panel shows $I_{\rm E}$ plotted against the displacement (in Euclid VIS pixel units) relative to the average motion of NGC\,6397. Probable cluster members are highlighted in black, while field stars are depicted in gray.}  
  \label{fig:ngc6397_rmap}
\end{figure*}

The resulting differential-reddening map derived from Euclid photomery is plotted in the left panel of Figure\,\ref{fig:ngc6397_rmap}. The middle panel of Figure\,\ref{fig:ngc6397_rmap} shows the differential-reddening corrected $I_{\rm E}$ vs.\,$I_{\rm E}-Y_{\rm E}$ CMD, whereas the displacements between the stellar positions in the Libralato's and Stetson's catalogs are plotted in the right panel against $I_{\rm E}$. This diagram has been used to separate the bulk of field stars (gray points) from the cluster members that we will use to study multiple populations over a wide field a view (black points). 

\begin{figure*} 
  \centering
 \includegraphics[height=8.6
 cm,trim={0.0cm 5.5cm 7.5cm 4.8cm},clip]{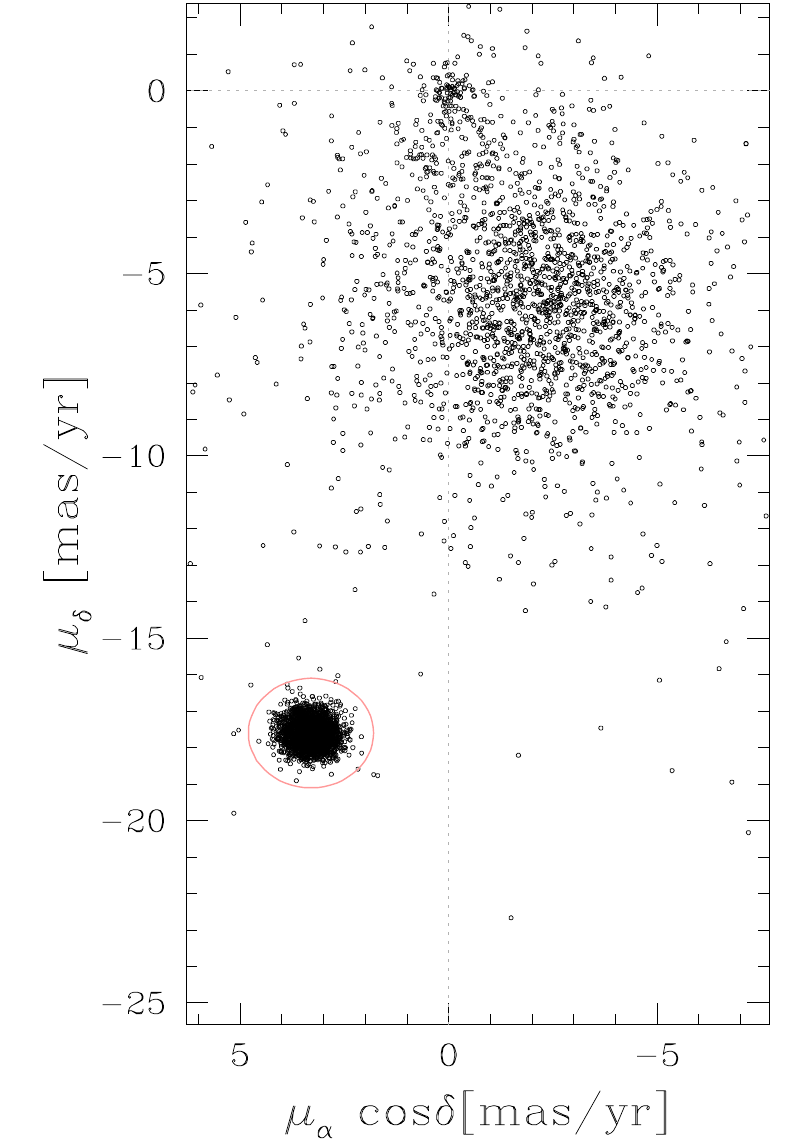}
 \includegraphics[height=8.5
 cm,trim={0.0cm 6.0cm 4.5cm 6.0cm},clip]{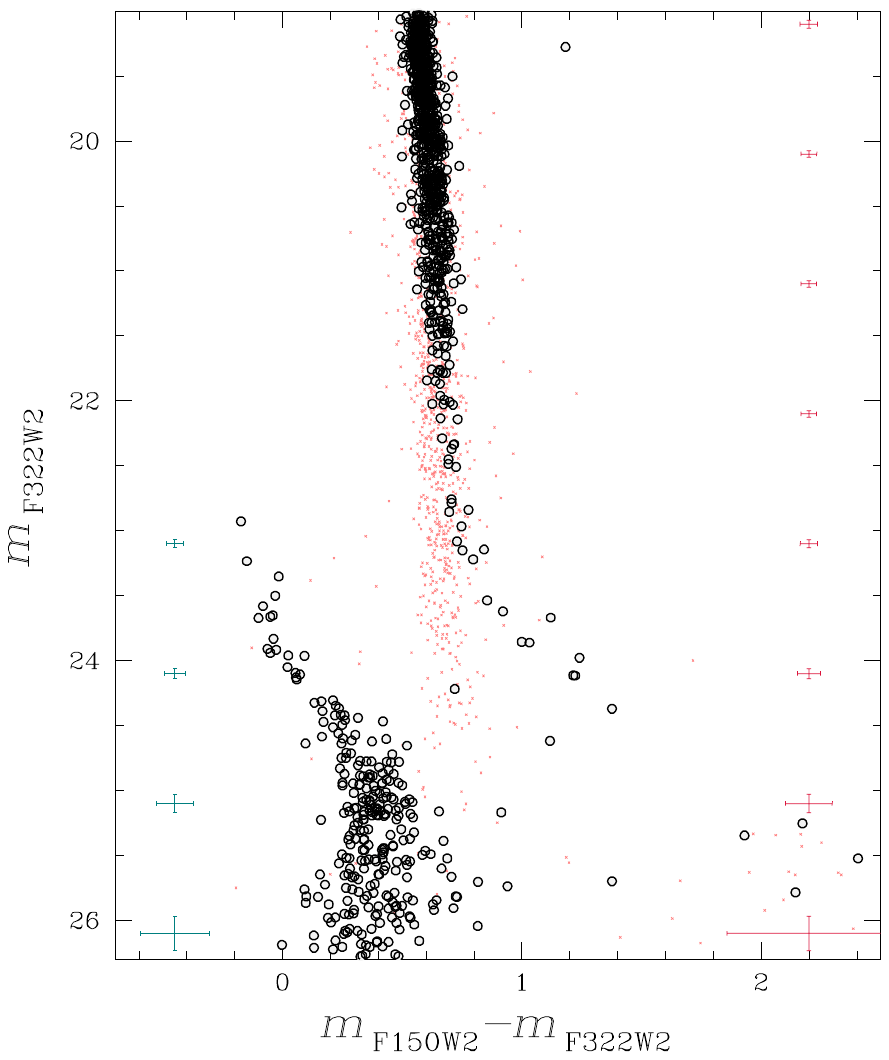}
  \caption{Proper-motion diagram of stars in the field A. The light-red circle encloses the bulk of cluster members (left). $m_{\rm F322W2}$ versus $m_{\rm F150W2}-m_{\rm F322W2}$ CMD from NIRCam data. Black circles and light-red points mark probable the proper-motion selected cluster members and field stars, respectively. The error bars colored teal and crimson represent the typical photometric uncertainties of white dwarfs and MS cluster stars, respectively, at different F322W2 magnitude values  (right). }  
  \label{fig:cmdJWST}
\end{figure*} 

The proper motions of stars in Field A are displayed in the left panel of Figure\,\ref{fig:cmdJWST}. As extensively discussed in previous studies on the kinematics of stars in the direction of NGC\,6397 \citep[e.g.][]{milone2006a, kalirai2007a}, this diagram reveals that cluster members, clustered around ($\mu_{\alpha} \cos \delta$, $\mu_{\delta}$) $\sim$ (3.4, $-$17.6) mas yr$^{-1}$, are well-separated from field stars, enabling robust identification of NGC\,6397 members. Additionally, the diagram highlights a broad overdensity of stars centered around ($\mu_{\alpha} \cos \delta$, $\mu_{\delta}$) $\sim$ ($-$2.5, $-$6.0) mas yr$^{-1}$, corresponding to stars from the Galactic bulge. A distinct cluster of sources near the origin of the reference frame is also apparent, consisting of background galaxies.

The proper motion diagrams derived from astrometry using Euclid, {\it HST}, and {\it JWST} data have been instrumental in disentangling field stars from cluster members and analyzing the internal kinematics of stellar populations. For instance, the right panel of Figure\,\ref{fig:cmdJWST} presents the $m_{\rm F322W2}$ vs.\,$m_{\rm F150W2}-m_{\rm F322W2}$ CMD from NIRCam photometry, where proper-motion-selected cluster members are marked as black points, while field stars are shown as light-red points.

\begin{figure*} 
  \centering
      \includegraphics[height=9.4cm,trim={1.0cm 4.5cm 0.0cm 4.5cm},clip]{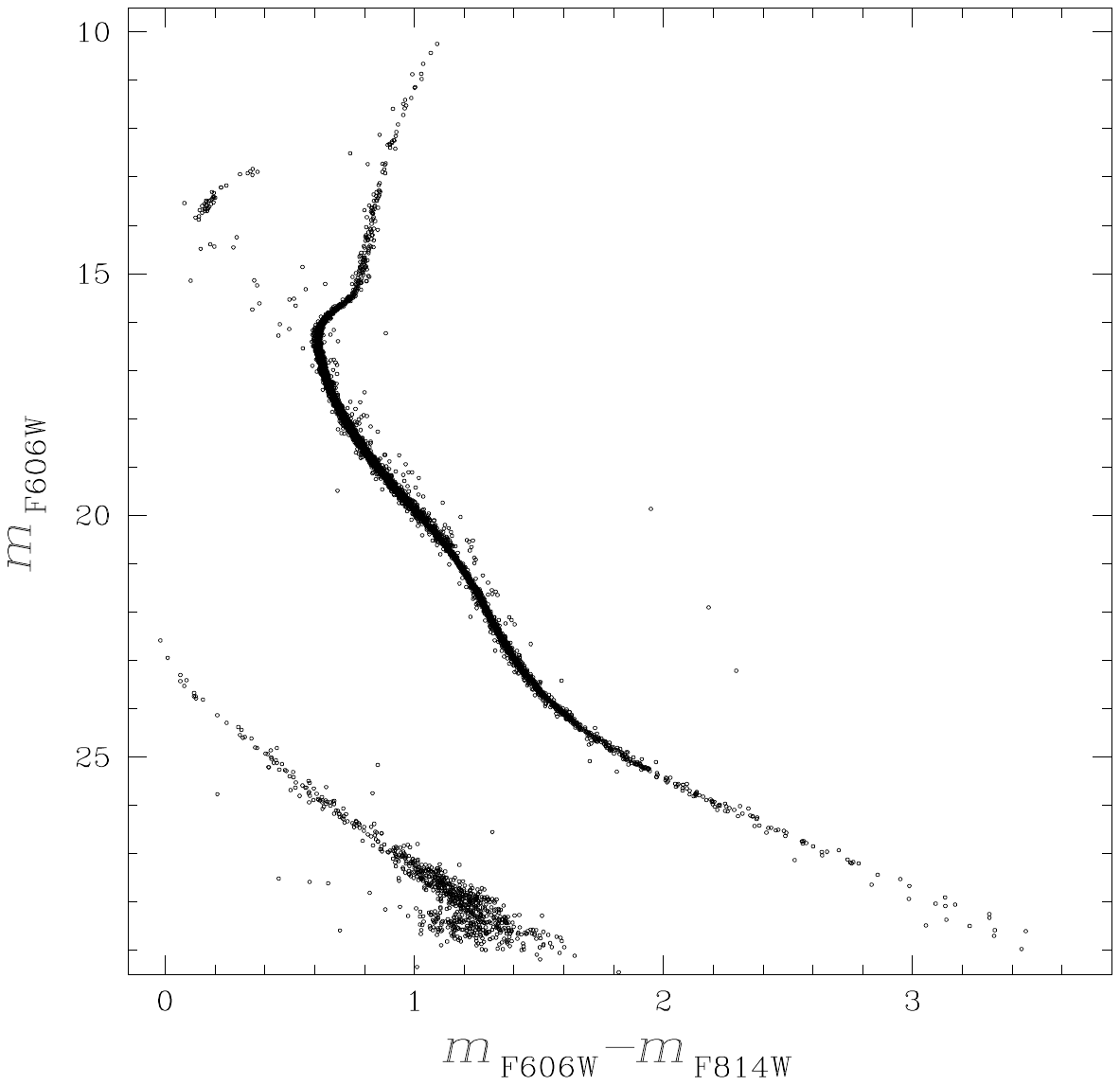}
   \includegraphics[height=9.4cm,trim={1.0cm 4.5cm 0.0cm 4.5cm},clip]{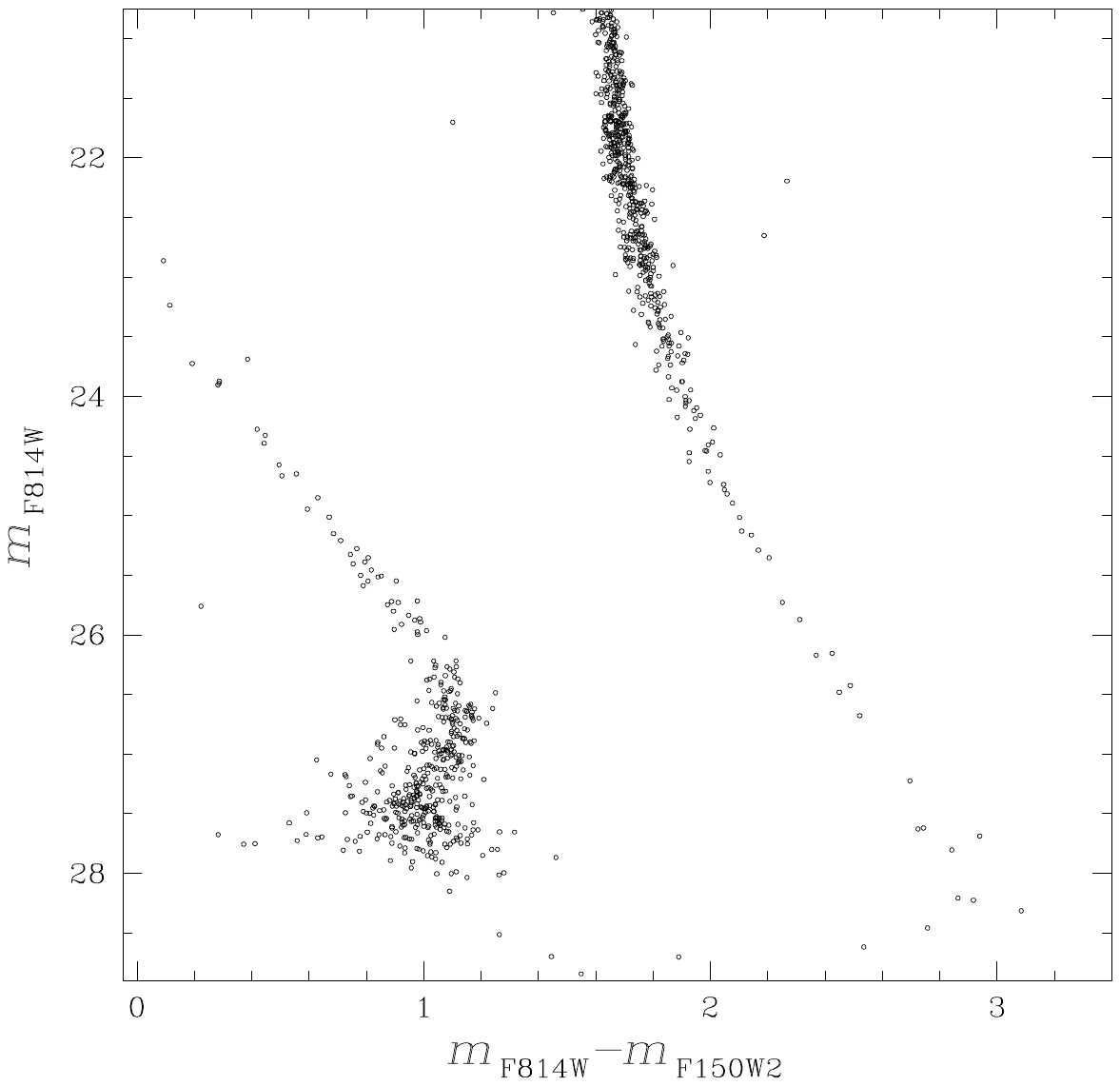}
  \caption{$m_{\rm F606W}$ vs.\,$m_{\rm F606W}-m_{\rm F814W}$ (left) and $m_{\rm F814W}$ vs.\,$m_{\rm F814W}-m_{\rm F150W2}$ (right) CMD of proper-motion selected cluster members.}  
  \label{fig:cmdJWSTHST}
\end{figure*} 

Figure\,\ref{fig:cmdJWSTHST} presents additional CMDs for proper-motion-selected cluster members. The left panel showcases the $m_{\rm F606W}$ vs.\,$m_{\rm F606W}-m_{\rm F814W}$ CMD for stars located in both the central field and Field A. This CMD covers a wide magnitude range of approximately 19 magnitudes and includes stars at various evolutionary stages. Specifically, it features the entire MS, extending from the turn-off point to the hydrogen-burning limit, as well as the sub-giant branch, the RGB, and the complete white-dwarf cooling sequence. This comprehensive CMD serves as a valuable tool for studying the proper motion distributions of stars across different evolutionary phases and different masses.

The right panel of Figure\,\ref{fig:cmdJWSTHST} illustrates the $m_{\rm F814W}$ vs.\,$m_{\rm F814W}-m_{\rm F322W2}$ CMD for proper-motion-selected faint stars of NGC\,6397 in Field A. The stellar sequence with colors exceeding $\sim 1.4$ represents the faintest part of the MS, extending down to the hydrogen-burning limit at $m_{\rm F814W} \sim 27$ mag, and includes the brightest brown dwarfs visible $m_{\rm F814W} \gtrsim 27$ mag. On the left, the white dwarf cooling sequence is prominently visible, showcasing notable slope changes around $m_{\rm F814W} \sim 26.5$ and 27.5 mag.
These CMDs, along with the one shown in Figure\,\ref{fig:cmdJWST}, will be instrumental in identifying multiple populations among M-dwarfs and analyzing their internal proper motions. For earlier studies of the faintest stars in NGC\,6397 based on similar data, we refer readers to \citet{richer2006a}, \citet{bedin2024a}, \citet{gerasimov2024a}, and references therein.

\section{Multiple populations with Euclid: a theoretical approach}\label{sec:theory}
The distinct chemical compositions of 2P stars in GCs, characterized by enhanced helium and nitrogen alongside reduced carbon and oxygen compared to 1P stars, drive the appearance of multiple stellar sequences in CMDs despite similar atmospheric parameters.

Here, we explore how variations in He, C, N, and O affect CMDs constructed with Euclid filters, applying methods previously used for other telescopes and filter sets \citep{sbordone2011a, dotter2015a, milone2018a, milone2023a, salaris2019a, li2022a}. 

We began by examining stellar populations characteristic of a GC with an intermediate metallicity of [Fe/H] = $-1.5$, extending our analysis to a metal-poor GC with [Fe/H] = $-1.9$ and a metal-rich GC with [Fe/H] = $-0.75$. For consistency, we used the same isochrones and synthetic spectra applied in previous studies of NIRCam photometry for M\,92 and 47\,Tucanae \citep{ziliotto2023a, milone2023a}.

Specifically, we selected helium-poor (Y = 0.246) and helium-rich (Y = 0.33) isochrones from the Dartmouth database \citep{dotter2008a}, both with an age of 13 Gyr and an $\alpha$-enhancement of [$\alpha$/Fe] = 0.4. At fifteen points along each isochrone, we extracted effective temperatures and surface gravities to model synthetic spectra representative of 1P and 2P stars. The 1P reference spectra, associated with Y = 0.246, assume solar carbon and nitrogen abundances with [O/Fe] = 0.4. In contrast, 2P spectra incorporate elemental variations with [C/Fe] = $-0.5$, [N/Fe] = 1.2, and [O/Fe] = $-0.1$. In addition,  we simulated spectra for stars with 2P-like C, N, and O abundances but Y=0.246, which allow us to visualize the effects of these light elements alone on stellar spectra, and spectra with Y=0.33 but 1P-like content of C, N, and O to explore the effects of helium changes alone on the spectra.

The model atmospheres are developed by \citet{milone2023a} and \citet{ziliotto2023a}  by using the ATLAS12 program \citep{kurucz1970a, kurucz1993a, sbordone2004a}, which leverages opacity-sampling and assumes a plane-parallel geometry in local thermodynamic equilibrium (LTE). These models span a range of temperatures (3500–6500 K) and surface gravities (log(g) = 2–5) with a fixed microturbulent velocity of 2 km/s. Molecular line data for species like C$_{2}$, CN, CO, H$_{2}$O, MgH, OH, SiH, 
   TiO, VO, ZrO, along with the latest atomic line lists, were included. Synthetic spectra were produced using SYNTHE \citep{kurucz1981a}, covering wavelengths from 1,000 to 51,000 \AA\, at high resolution. Magnitudes were calculated by integrating these spectra over the {\it HST}, {\it JWST} and Euclid bandpasses, and 2P magnitudes were derived by applying differential offsets to the 1P isochrone values.

\begin{figure*} 
  \centering
 \includegraphics[height=9.5cm,trim={0.0cm 5.0cm 0.0cm 2.0cm},clip]{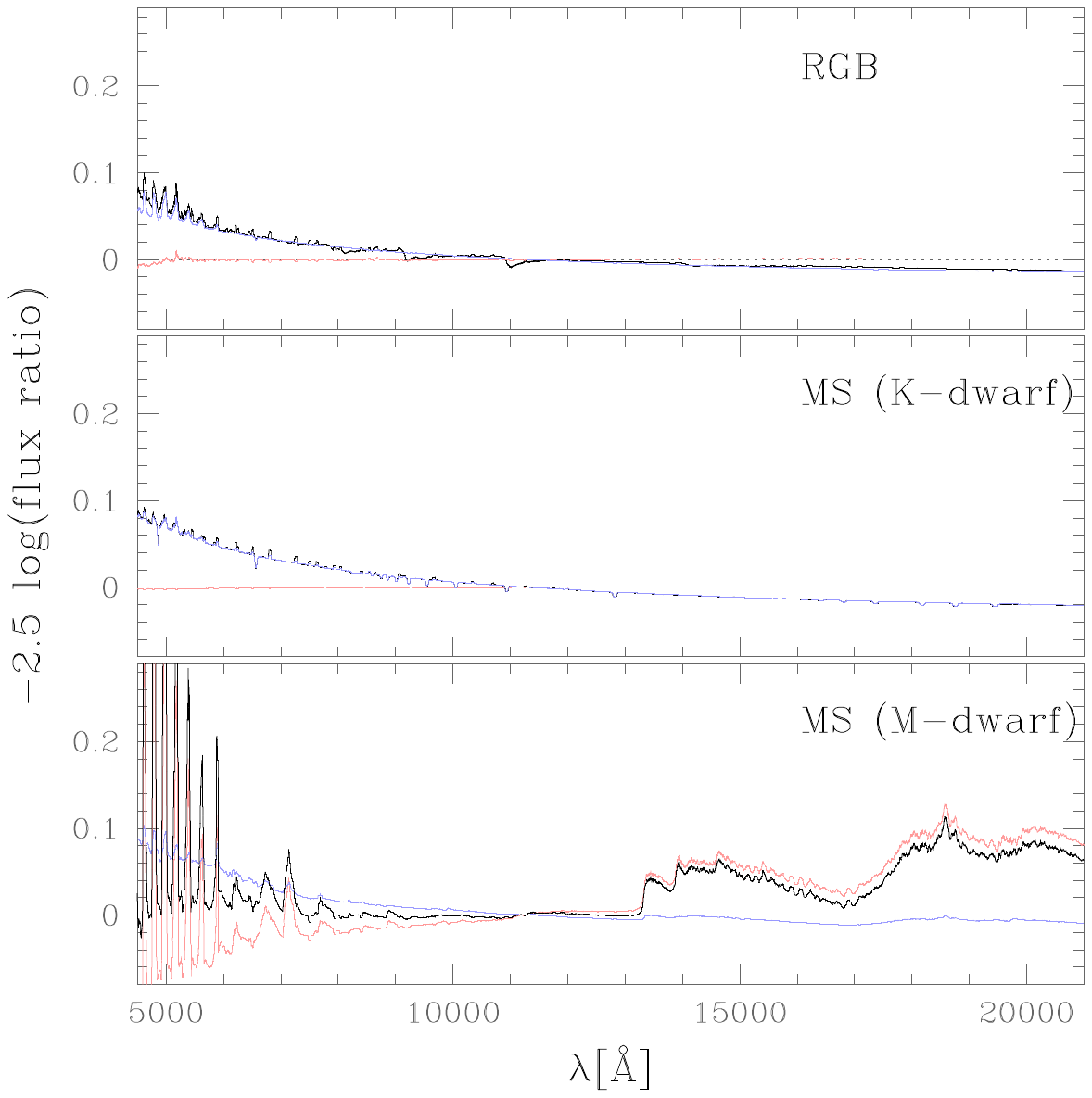}
 \includegraphics[height=9.5cm,trim={0.0cm 5.0cm 0.0cm 2.0cm},clip]{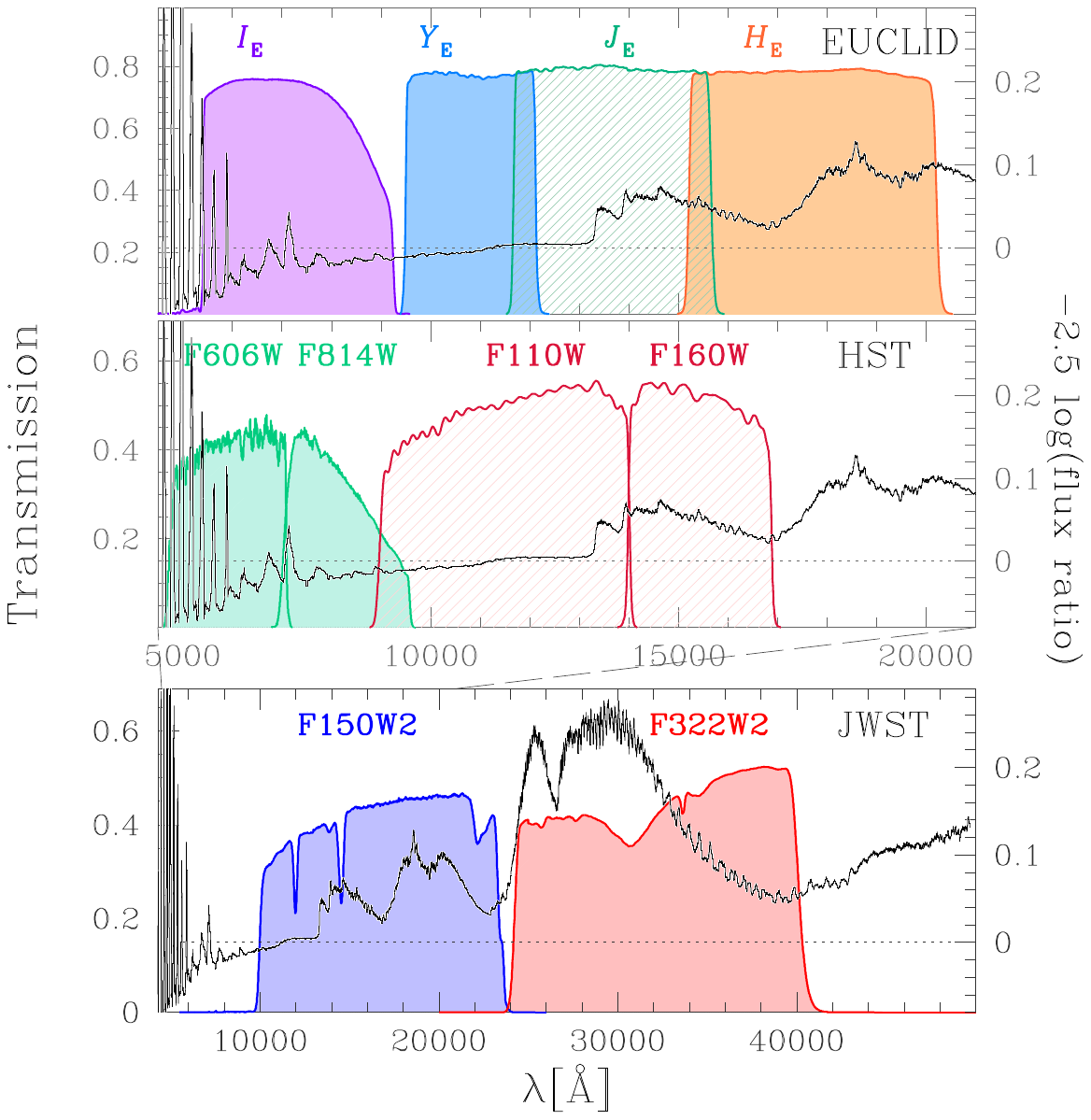}
  \caption{\textit{Left panels}.
  Flux ratios were calculated for simulated spectra of stars with identical luminosity in the NIRCam F115W band but different chemical compositions. The 2P spectra are helium and nitrogen-enhanced while being carbon and oxygen-depleted compared to the 1P spectra. Light-blue spectra represent stars with the same C, N, O abundances as 1P stars but enriched in helium, while light-red spectra correspond to stars with 1P-like helium content but  different C, N, and O abundances.  
  The bottom, middle, and top panels correspond to M-dwarfs, K-dwarfs, and RGB stars, respectively \citep{milone2023a}. 
  For 1P stars and 2P-like stars with Y=0.246, 
  the effective temperatures, T$_{\rm eff}$, and gravities, log(g), 
  are 3762\,K, 5901\,K, 4746\,K 
  and 5.03, 4.50, 1.82, respectively. Helium-rich stars were modeled with 
  T$_{\rm eff}$=3808\,K, 6048\,K, 4803\,K, 
 and log(g)=5.03, 4.48, 1.78, based on \cite{dotter2008a} isochrones. \textit{Right panels.} Transmission curves of the filters used in this paper. The Euclid filters (top), the F606W and F814W ACS/WFC and F110W F160W IR/WFC3 filters on board {\it HST} (middle), and the F150W2 and F322W2 filters of NIRCam/JWST. We superimpose on each panel the flux ratio between 2P and 1P M-dwarf with different helium and C, N, O abundances.  }
  \label{fig:spettri}
\end{figure*}

The left panels of Figure\,\ref{fig:spettri} illustrate the relative fluxes of simulated spectra for an M-dwarf (bottom), a K-dwarf (middle), and an RGB star (top), compared to the 1P spectrum with the same F115W magnitude. Unlike \cite{milone2023a}, which explored multiple populations' effects on NIRCam photometry, this study focuses on the wavelength range observed by Euclid.

As noted in earlier studies, variations in C, N, and O have minimal impact on RGB and MS stars brighter than the MS knee \citep[e.g.,][]{milone2012b}. Helium variations mostly affect stellar structure and have negligible effect on stellar atmospheres. Helium-rich stars, being hotter than their helium-normal counterparts with the same F115W luminosity, display brighter optical magnitudes, as shown in the top and middle panels \citep{dantona2002a, norris2004a, sbordone2011a}.

The bottom panel highlights a distinct feature of M-dwarfs: beyond $\sim$13,000 \AA\, O-rich stars exhibit stronger absorption than O-poor stars due to oxygen-based molecules like water vapor. These absorption features overlap with the spectral regions covered by Euclid's J$_{\rm E}$ and H$_{\rm E}$ filters as shown in the top-right panel of Figure\,\ref{fig:spettri}, where we compare the filter throughputs with the flux ratio of 2P to 1P stars.  This figure also reveals that in the optical, molecular absorption compensates the helium effects, resulting in similar spectra below $\sim$7,500 \AA.

A visual comparison of flux ratio plots between 1P and 2P stars with identical F115W magnitudes, as a function of wavelength, reveals that stellar populations with [Fe/H]=$-$0.75 and [Fe/H]=$-$2.3 display comparable qualitative trends \citep{milone2023a, ziliotto2023a}. However, a notable exception arises from the strong molecular absorption features affecting optical wavelengths in the spectra of M-dwarfs with [Fe/H]=$-$0.75.

\begin{figure*} 
  \centering
 \includegraphics[height=9.5cm,trim={0.5cm 5.8cm 9.75cm 3.5cm},clip]{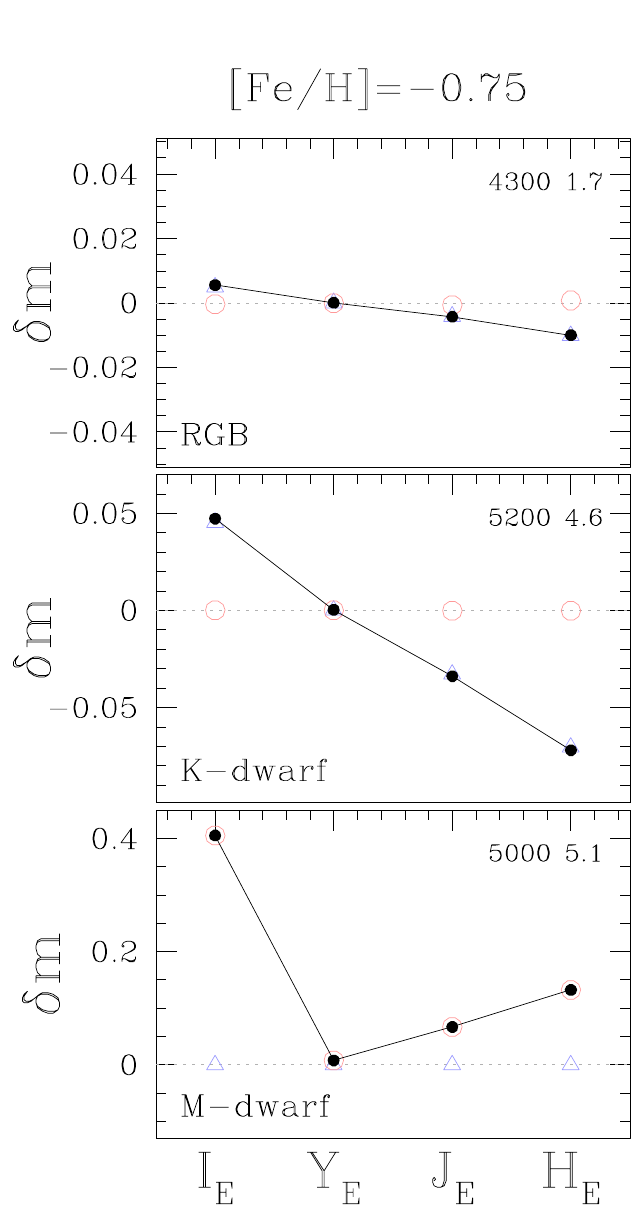}
 \includegraphics[height=9.5cm,trim={1cm 5.8cm 9.75cm 3.5cm},clip]{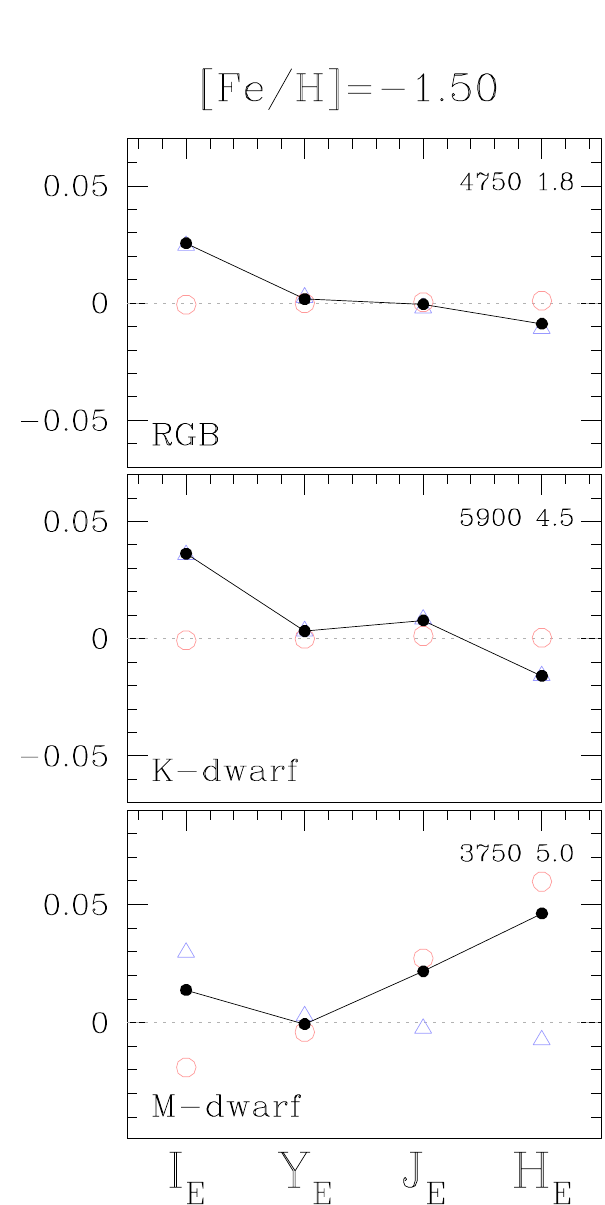}
 \includegraphics[height=9.5cm,trim={1cm 5.8cm 9.75cm 3.5cm},clip]{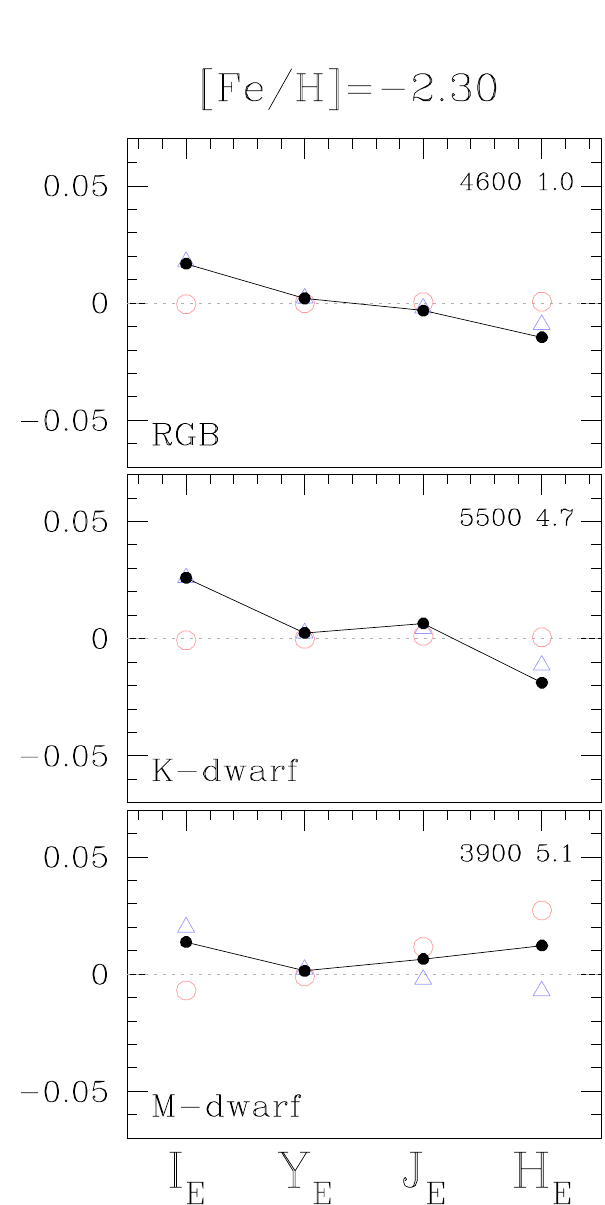}
  \caption{Magnitude difference between a 1P star and a star with the same F115W magnitude and a chemical composition that resembles 2P stars.  The azure points correspond to stars with the same chemical composition as 1P stars but enhanced in helium (Y=0.33). The light-red points are derived from spectra that share the same helium content as 1P stars but are enhanced in  N and depleted  in  both C and O. Left, middle, and right panels show spectra with [Fe/H]=$-$0.75, $-$1.50, and $-$2.30 \citep{milone2023a, ziliotto2023a}, respectively. The results for RGB, bright-MS stars, and M -dwarfs are shown in the top, middle, and bottom panel respectively. We display in the top-right corner of -each panel the effective temperature and the logarithm of gravity for 1P spectra.}  \label{fig:DCOL}
\end{figure*}

Figure\,\ref{fig:DCOL} quantifies these behaviors by showing the magnitude differences, $\delta_{\rm m}$ (where $m=I_{\rm E}, Y_{\rm E}, J_{\rm E},$ and $H_{\rm E}$), for the pairs of RGB, K-dwarfs, and M-dwarfs analyzed in Figure\,\ref{fig:spettri} and earlier works \citep{milone2023a, ziliotto2023a}. Each pair consists of stars with identical F115W magnitudes but differing helium and C, N, O abundances. Notably, these are the same simulated spectra pairs previously used by \citet{milone2023a} and \citet{ziliotto2023a} to evaluate magnitude differences in UVIS/WFC3, IR/WFC3, and NIRCam data, thus allowing direct comparison of the results from Euclid, {\it HST}, and {\it JWST} photometry.

\begin{figure*} 
  \centering
 \includegraphics[height=7cm,trim={0cm 5.1cm 4.8cm 4cm},clip]{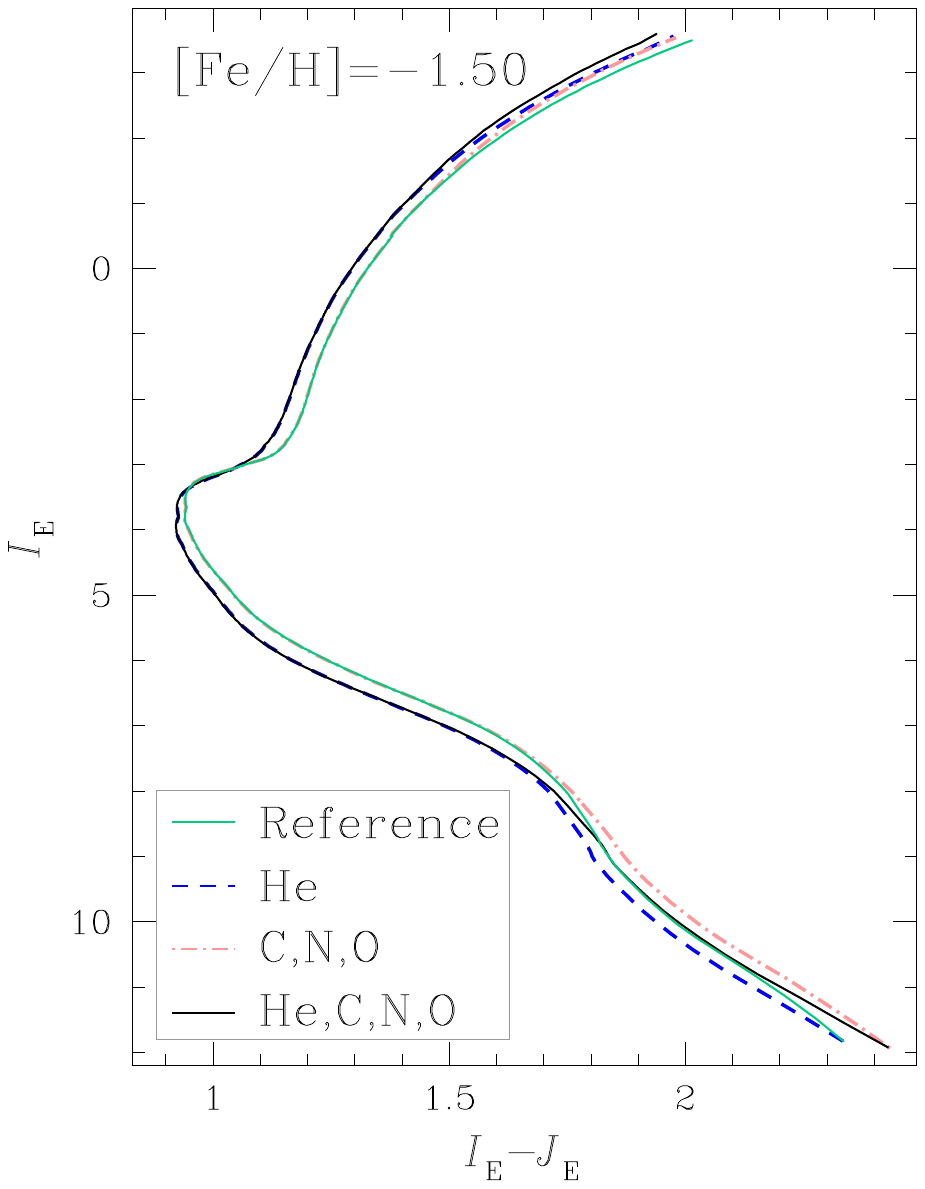}
 \includegraphics[height=7cm,trim={0cm 5.1cm 4.8cm 4cm},clip]{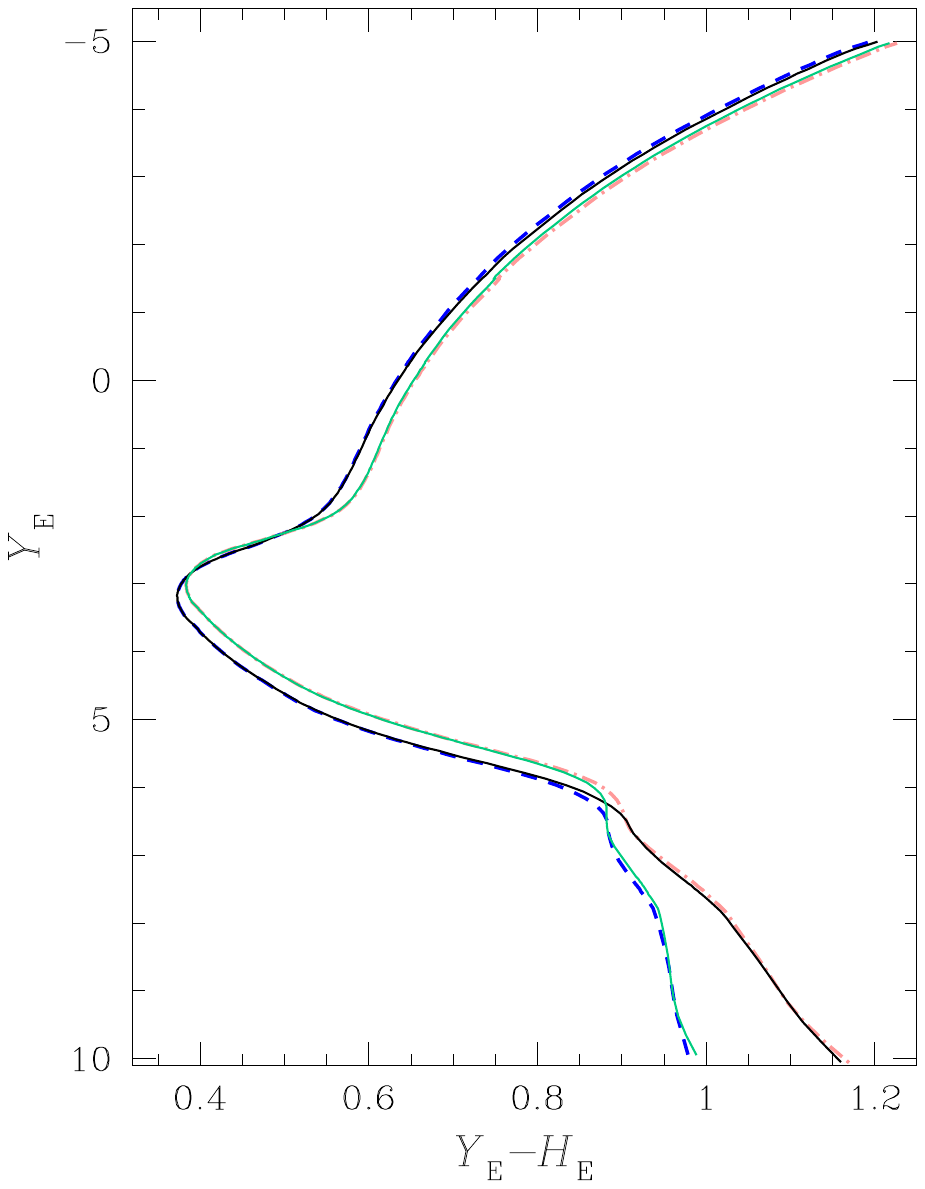}
 \includegraphics[height=7cm,trim={0cm 5.1cm 4.8cm 4cm},clip]{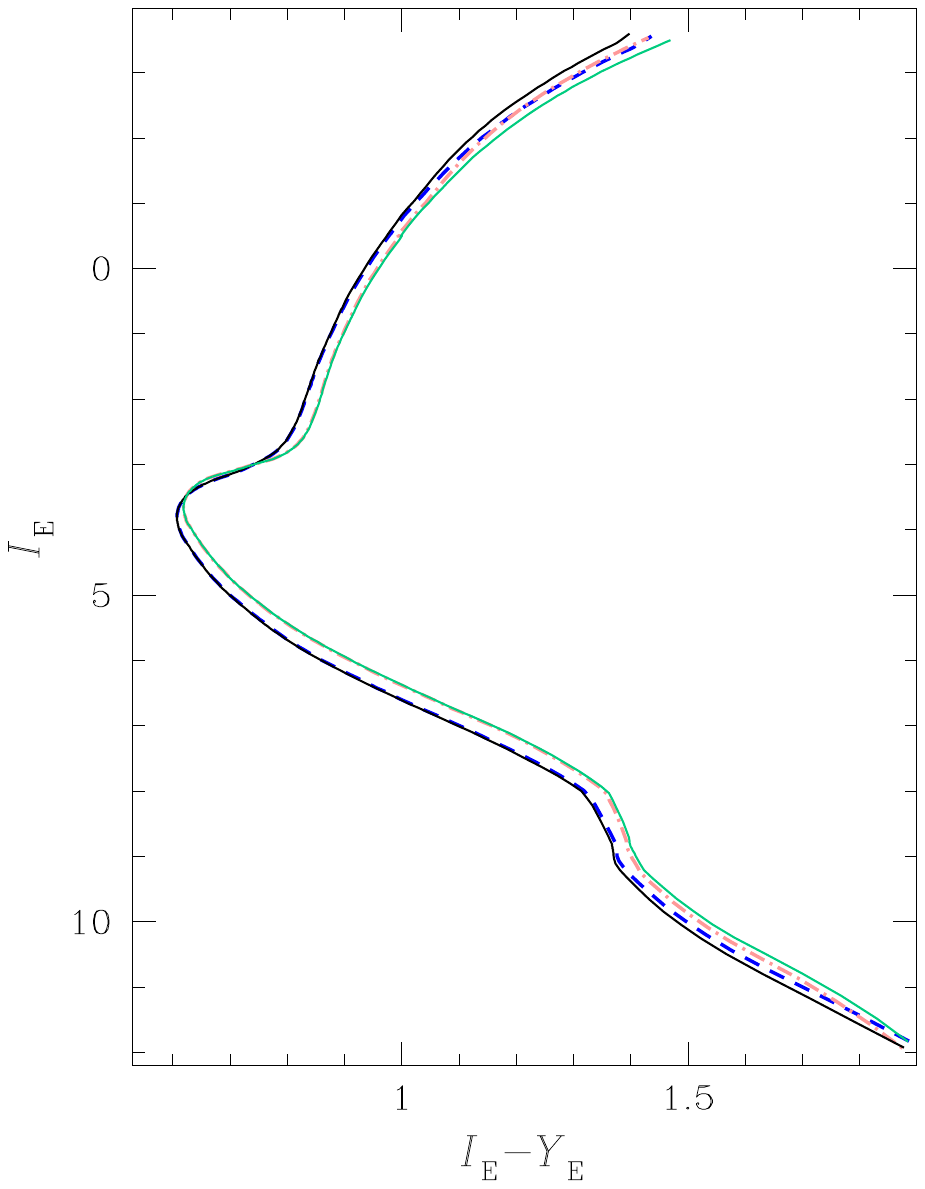}

\includegraphics[height=7cm,trim={0cm 5.1cm 4.8cm 4cm},clip]{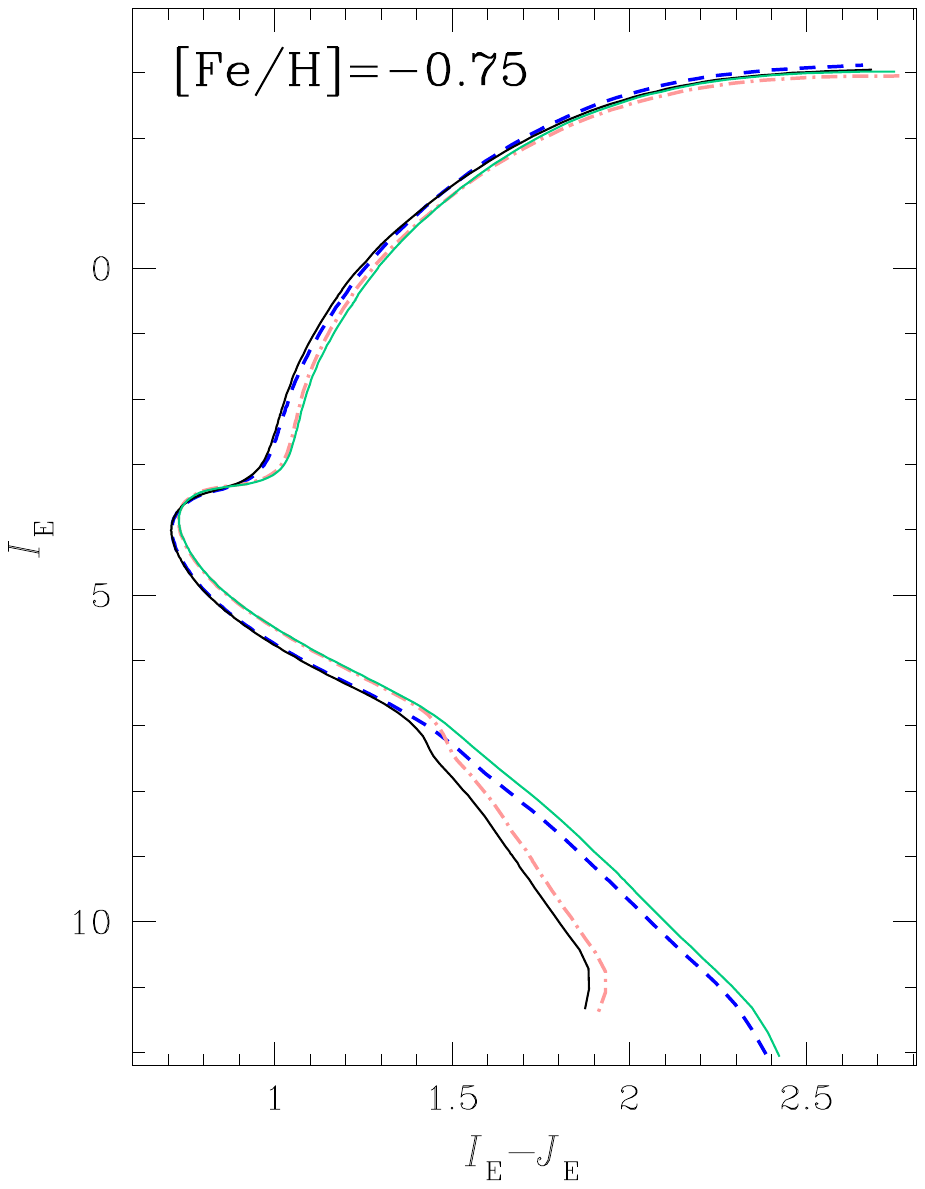}
 \includegraphics[height=7cm,trim={0cm 5.1cm 4.8cm 4cm},clip]{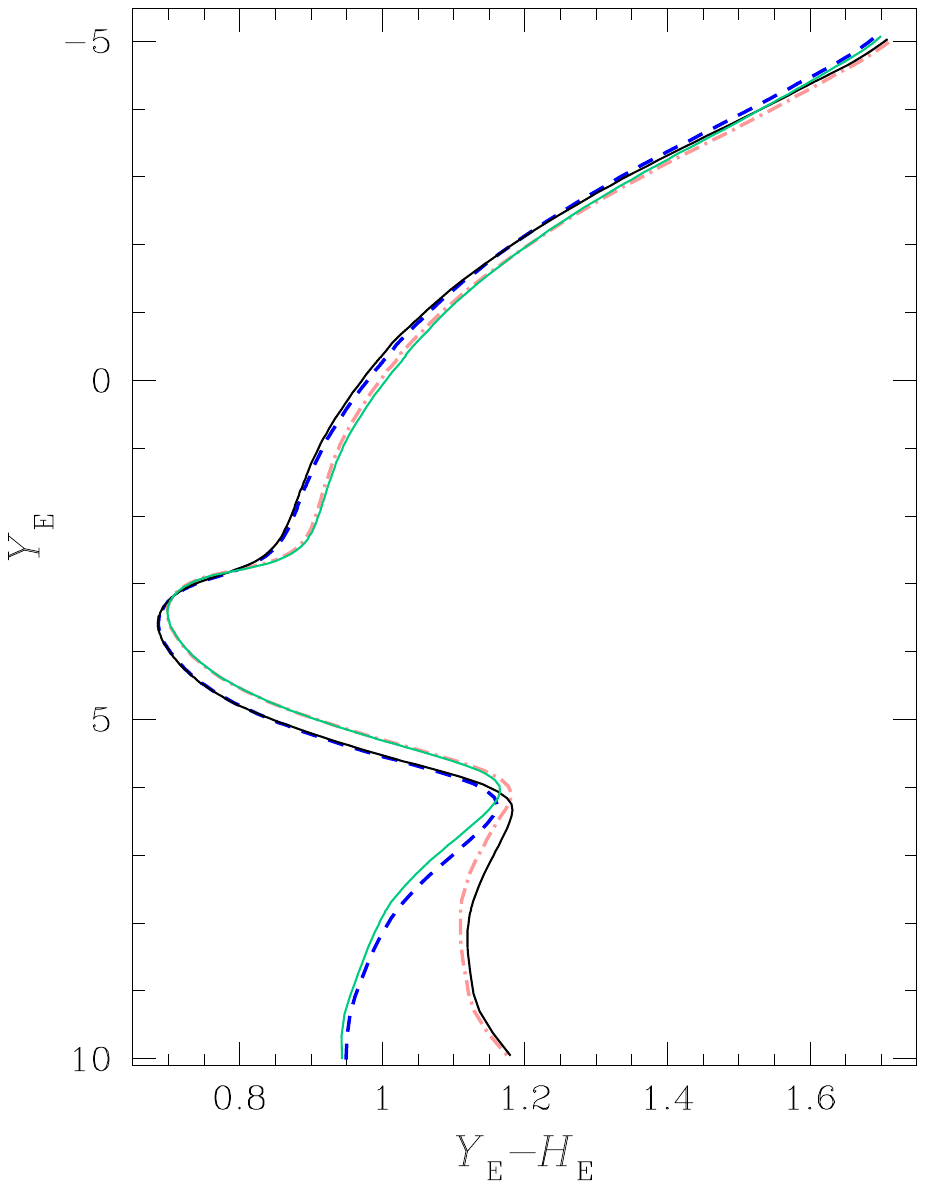}
 \includegraphics[height=7cm,trim={0cm 5.1cm 4.8cm 4cm},clip]{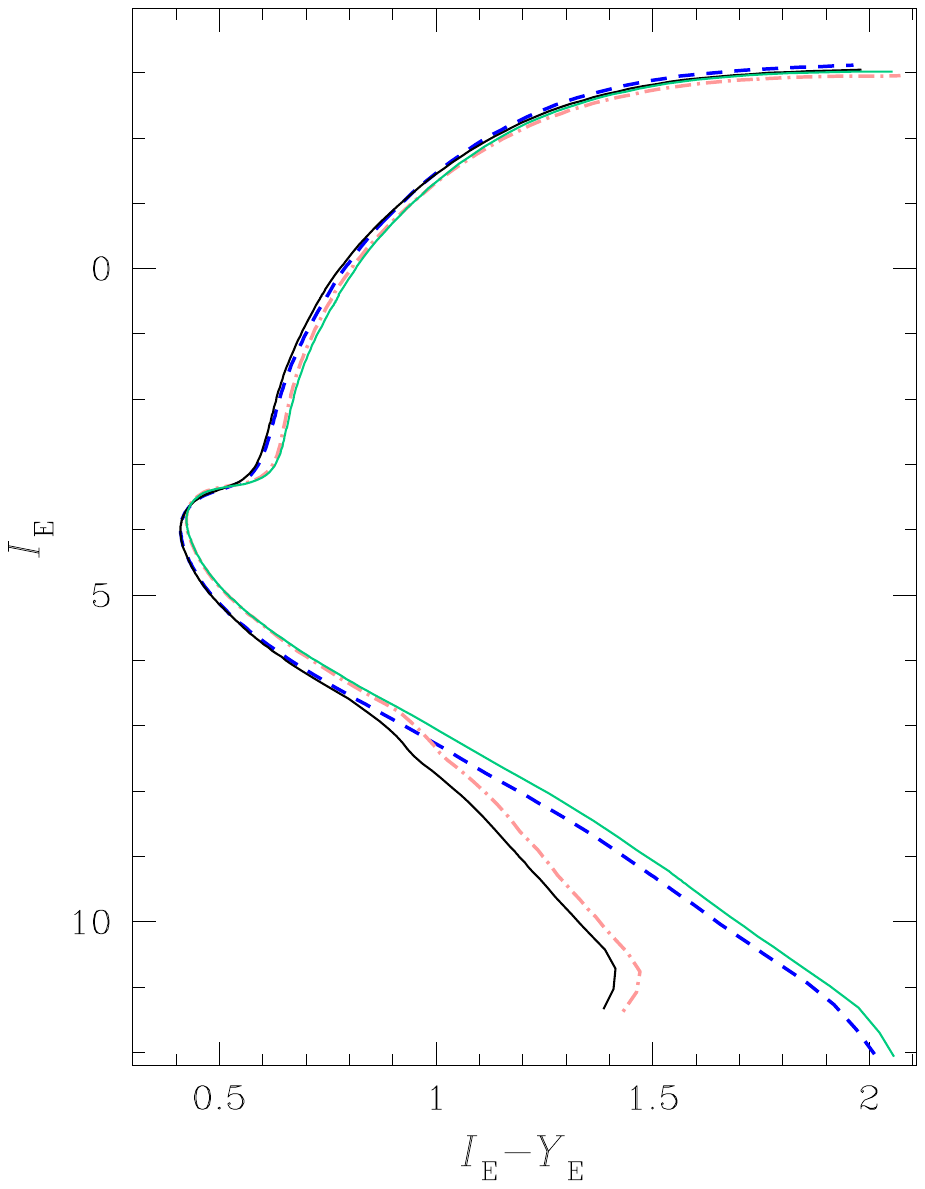}

  \caption{13 Gyr isochrones from the Dartmouth database \citep{dotter2008a} with [$\alpha$/Fe]=0.4 and  with [Fe/H]=$-$1.50 (top panels) and [Fe/H]=$-0.75$ (bottom panels). Isochrones with pristine helium content (Y=0.246 and Y=0.254 for the top- and bottom-panels, respectively) are shown in aqua and light red, while those enriched in helium (Y=0.33) are displayed in blue and black. The light-red and black isochrones have chemical compositions with carbon and oxygen depletion of 0.5 and 0.9 dex, respectively, along with a nitrogen enhancement of 1.2 dex. In contrast, the aqua and blue isochrones represent stars with [O/Fe]=0.4 and solar abundances of carbon and nitrogen.}  
  \label{fig:ISO}
\end{figure*}

The simulated spectra were used to calculate the colors and magnitudes of the isochrones depicted in Figure\,\ref{fig:ISO}. For [Fe/H] = $-$1.5, helium-rich isochrones display bluer $I_{\rm E}-J_{\rm E}$ colors compared to the aqua and light-red isochrones (Y = 0.246) that share the same C, N, and O abundances as the blue and black isochrones, respectively (top-left panel of Figure\,\ref{fig:ISO}). A notable exception occurs in the SGB region, where isochrones with different helium content become intertwined.
Above the MS knee, the aqua isochrones, which mimic 1P stars and are taken as references, and light-red isochrones are nearly indistinguishable. However, below the MS knee, they diverge, with O-poor isochrones appearing bluer than those with [O/Fe] = 0.4.
Similarly, the blue and black isochrones, both helium-enhanced (Y = 0.33) but with different C, N, and O abundances, exhibit comparable trends. Interestingly, in the $I_{\rm E}$ vs.\,$I_{\rm E}-J_{\rm E}$ CMD, helium enhancement and oxygen depletion have opposing effects on stellar colors. 

Above the MS knee, the isochrones with [Fe/H] = $-$1.5 and different He, C, N, and O abundances exhibit comparable trends in both the $I_{\rm E}$ vs.\,$I_{\rm E}-J_{\rm E}$ and $Y_{\rm E}$ vs.\,$Y_{\rm E}-H_{\rm E}$ CMDs. Below the MS knee, however, the pairs of isochrones with identical C, N, and O abundances but different helium content overlap almost entirely in the $Y_{\rm E}$ vs.\,$Y_{\rm E}-H_{\rm E}$, while oxygen-poor isochrones become distinctly redder than those with [O/Fe] = 0.4, due to the increasing influence of molecular absorption bands in the cooler atmospheres of M-dwarfs.

The MS of isochrones with different C, N, and O abundances but identical helium content is nearly indistinguishable in the $I_{\rm E}$ vs.\,$I_{\rm E}-Y_{\rm E}$ CMD (left panel of Figure\,\ref{fig:ISO}). This scarce sensitivity to abundance variations in clusters with [Fe/H]=$-$1.5 and moderate helium differences makes this diagram particularly effective for identifying binaries composed of MS stars throughout the entire MS. Moreover, it enables precise determination of the MS knee position, which is crucial for accurate age estimates based on the relative locations of the MS knee and the MS turn-off.

The isochrones with [Fe/H]=$-$2.30 exhibit qualitatively similar behavior to those with [Fe/H]=$-$1.50 in the CMDs shown in Figure\,\ref{fig:ISO}, albeit with smaller color separations. In contrast, metal-rich isochrones ([Fe/H]=$-$0.75) display notable differences compared to those with [Fe/H]=$-$1.50 or $-$2.30, particularly below the MS knee. These discrepancies arise primarily from the stronger molecular band absorption in the optical region of M-dwarf spectra \citep[e.g.][see their figure 12]{milone2023a}. 
In particular, as shown in the bottom panels of Figure\,\ref{fig:ISO}, the isochrones with different oxygen abundances have different colors below the MS-knee, with the O-poor isochrones drawing sequences with redder $I_{\rm E}-J_{\rm E}$ and $I_{\rm E}-Y_{\rm E}$ colors and bluer $Y_{\rm E}-H_{\rm E}$ colors than the O-rich isochrones. We verified that these conclusion, obtained from isochrones based on \citet{dotter2008a} models are confirmed when using isochrones from the BaSTI database \citep{pietrinferni2021a}.

To further explore multiple stellar populations in photometric diagrams derived from Euclid photometry, we utilized isochrones with chemical compositions representative of 1P and 2P stars. We simulated six distinct isochrones with different levels of He, C, N, O, and Fe, as detailed in Table\,\ref{tab:iso6}. Isochrones I1 are characteristic of 1P stars, while I2, I3, and I4 mimic 2P stars, maintaining the same iron abundance as I1 but incorporating different degrees of light-element variations. Isochrones I5 mirror the C, N, O, and Fe content of I1 but include enhanced helium abundances. Finally, I6 isochrones have the same helium and light-element abundances relative to iron as I1 but exhibit an increased [Fe/H] by 0.1 dex, accounting for star-to-star metallicity differences that drive the extended 1P sequence observed in chromosome maps of numerous clusters \citep{milone2015a, marino2019b, legnardi2022a, legnardi2024a, lagioia2024a}. 

\begin{figure*} 
  \centering
  \includegraphics[height=9cm,trim={0cm 5.1cm 0.0cm 11.5cm},clip]{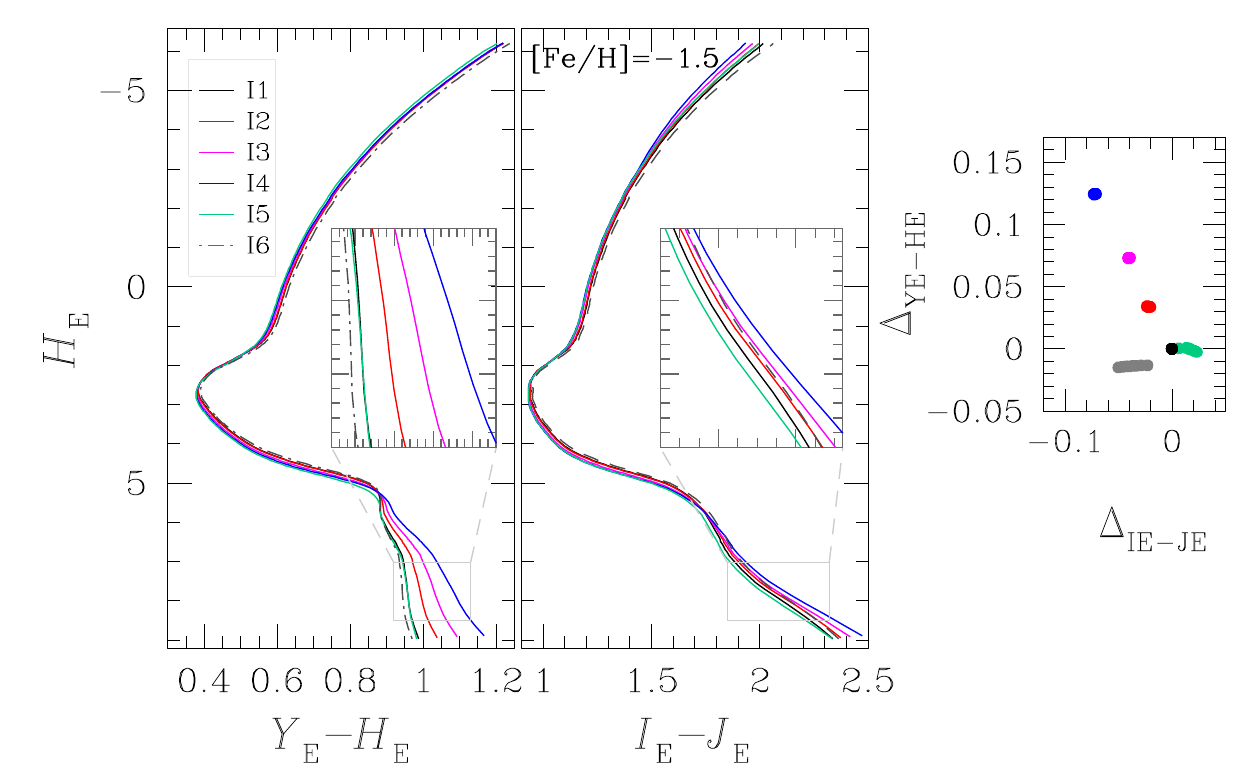}
  \includegraphics[height=9cm,trim={0cm 5.1cm 0.0cm 11.5cm},clip]{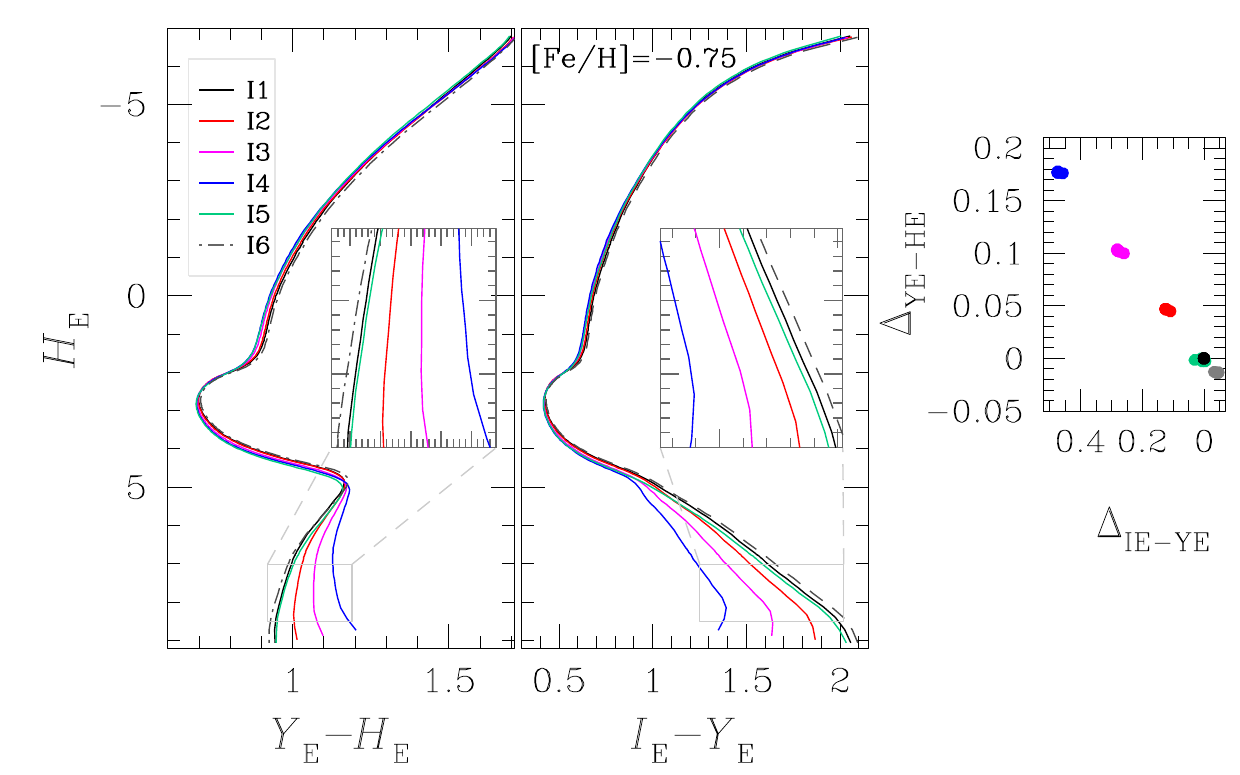}
  \caption{\textit{Top panels.} Isochrones with different contents of helium, carbon, nitrogen, and oxygen in the $H_{\rm E}$ vs.\,$Y_{\rm E}-H_{\rm E}$ (left) and $H_{\rm E}$ vs.\,$I_{\rm E}-J_{\rm E}$ (right) CMDs. The right panel shows the ChM for the M-dwarfs highlighted in the insets of the left and middle panels.
  All isochrones in the top panels have [Fe/H]=$-$1.5 with the exception of the isochrones I6, which are enhanced in iron abundance by 0.1 dex with respect to the other isochrones. Bottom panels represent the $H_{\rm E}$ vs.\,$Y_{\rm E}-H_{\rm E}$ (left) and $H_{\rm E}$ vs.\,$I_{\rm E}-Y_{\rm E}$ (right) CMDs, and the ChM derived from these CMDs for the M-dwarfs plotted in the CMD insets (right). The isochrones I1-I5 in the bottom panels share the same iron to hydrogen ratios,  [Fe/H]=$-0.75$, while isochrones I6 have [Fe/H]=$-0.65$.}
  \label{fig:ChMteo}
\end{figure*}

The $H_{\rm E}$ vs.\,$Y_{\rm E}-H_{\rm E}$ and $H_{\rm E}$ vs.\,$I_{\rm E}-J_{\rm E}$ CMDs, displayed in the top panels of Figure\,\ref{fig:ChMteo}, offer the most effective separation of stellar populations with varying C, N, O abundances and [Fe/H]=$-$1.5 below the MS knee. Consequently, the $\Delta_{\rm YE-HE}$ vs.\,$\Delta_{\rm IE-JE}$ chromosome map (ChM) optimally distinguishes multiple populations with different light-element compositions in GCs. The clustering of I6 stars near I1 stars further indicates that this ChM is poorly sensitive to the small metallicity variations typically found among 1P stars in Type\,I GCs. The analysis of isochrones with [Fe/H]=$-$2.3 and the same abundances of helium, carbon, nitrogen and oxygen as those shown in the top panels of Figure\,\ref{fig:ChMteo} provide the same conclusions.

\begin{table}
  \caption{Chemical compositions of the isochrones I1--I6.}
\begin{tabular}{l c c c c l }
\hline \hline
 ID & Y & [C/Fe]  & [N/Fe] & [O/Fe] & [Fe/H]  \\
 I1 & 0.246 &    0.00    &  0.00  &     0.40  & $-$1.50  \\
 I2 & 0.252 & $-$0.05    &  0.70  &     0.20  & $-$1.50  \\
 I3 & 0.276 & $-$0.25    &  1.00  &  $-$0.05  & $-$1.50 \\
 I4 & 0.292 & $-$0.35    &  1.20  &  $-$0.30  & $-$1.50  \\
 I5 & 0.292 &    0.00    &  0.00  &     0.40  & $-$1.50  \\
 I6 & 0.246 &    0.00    &  0.00  &     0.40  & $-$1.40  \\
\hline
I1 & 0.254 &    0.00    &  0.00  &     0.40  & $-$0.75  \\
 I2 & 0.261 & $-$0.05    &  0.70  &     0.20  & $-$0.75  \\
 I3 & 0.284 & $-$0.25    &  1.00  &  $-$0.05  & $-$0.75  \\
 I4 & 0.300 & $-$0.35    &  1.20  &  $-$0.30  & $-$0.75  \\
 I5 & 0.300 &    0.00    &  0.00  &     0.40  & $-$0.75  \\
 I6 & 0.254 &    0.00    &  0.00  &     0.40  & $-$0.65  \\
     \hline\hline
\end{tabular}
  \label{tab:iso6}
 \end{table}

The bottom panels of Figure\,\ref{fig:ChMteo} display isochrones with [Fe/H] = $-$0.75. As shown in the left panel, the relative behavior of the metal-rich I1–I6 isochrones is similar to what is observed for lower metallicities ([Fe/H] = $-$1.5 and $-$2.3). However, the rise of molecular absorption features, particularly in the $I_{\rm E}$ band for metal-rich stars fainter than the MS knee, makes colors involving this filter highly sensitive to variations in oxygen abundance. Specifically, 2P stars, which are oxygen-poor, experience less molecular absorption and thus appear brighter in the $I_{\rm E}$ band compared to 1P stars with higher oxygen content.

Among the investigated colors, the $I_{\rm E}-Y_{\rm E}$ color, used to construct the CMD in the bottom-middle panel of Figure\,\ref{fig:ChMteo}, provides the most pronounced separation between the multiple populations of M dwarfs. While $I_{\rm E}-J_{\rm E}$ and $I_{\rm E}-H_{\rm E}$ colors are also effective at distinguishing 1P and 2P stars, the molecular absorption effects on the $J_{\rm E}$ and $H_{\rm E}$ magnitudes partially counteract the absorption seen in $I_{\rm E}$. Interestingly, at [Fe/H] = $-$0.75, O-rich stars exhibit bluer $I_{\rm E}-J_{\rm E}$ colors than O-poor stars—a trend that is reversed in the simulated CMDs for lower metallicities.

Pseudo-colors, such as those derived from $C_{\rm IE,YE,JE}=$($m_{\rm IE}-m_{\rm YE}$)$-$($m_{\rm YE}-m_{\rm HE}$) or the $\Delta_{\rm YE-HE}$ vs.\,$\Delta_{\rm IE-YE}$ ChM plotted in the bottom-right panel of Figure\,\ref{fig:ChMteo}, offer an even greater ability to disentangle multiple populations and provide a valuable tool for studying their properties.

\begin{figure*} 
  \centering
 \includegraphics[height=8cm,trim={0.5cm 5.0cm 0.0cm 9.5cm},clip]{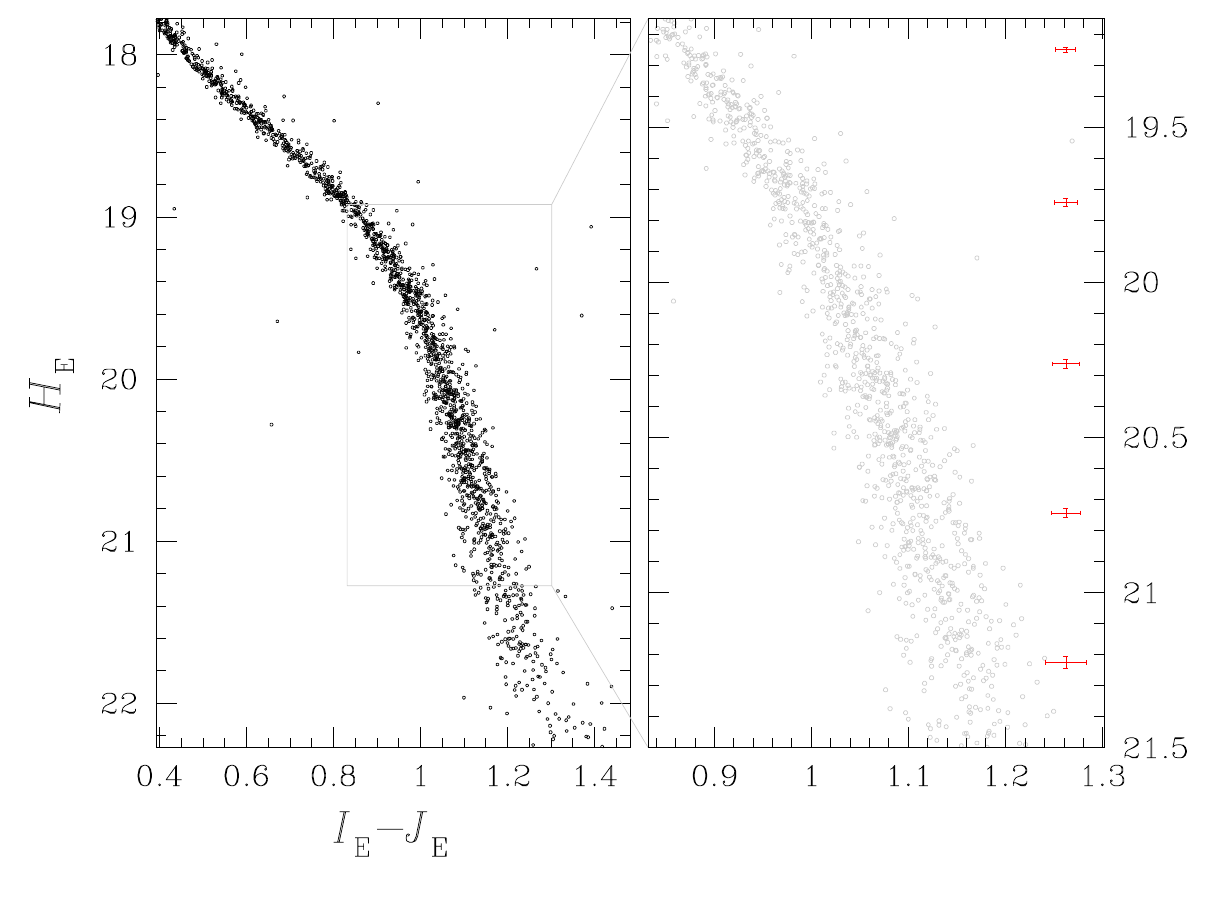}
  \includegraphics[height=8cm,trim={0.5cm 5.0cm 6.5cm 9.5cm},clip]{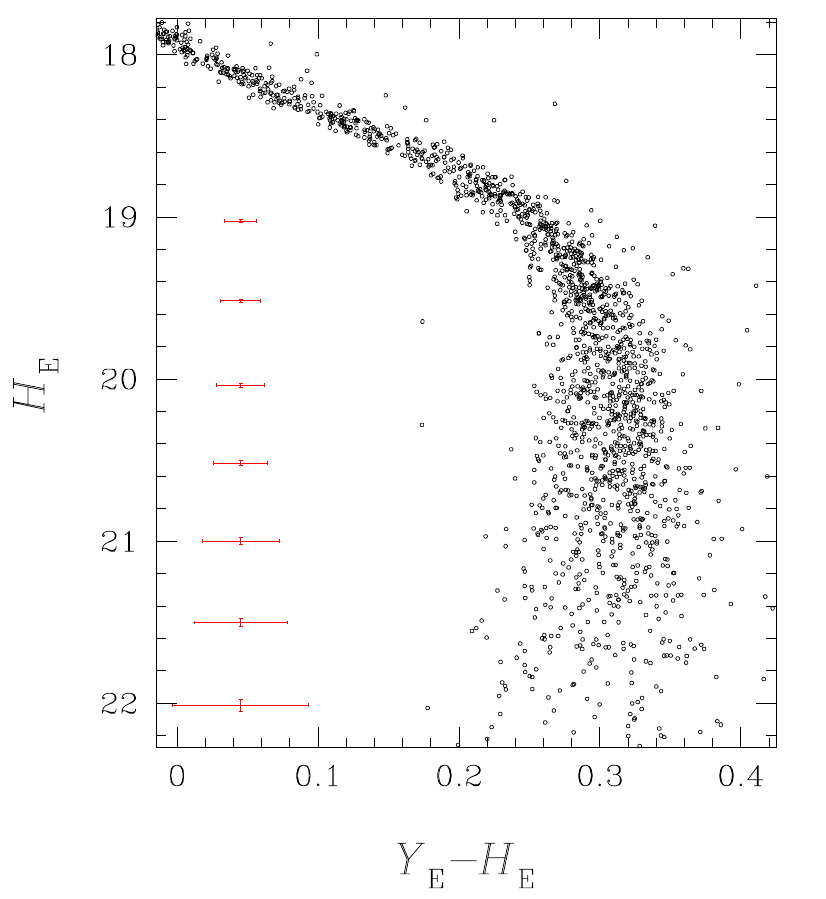}
  \caption{CMDs of proper-motion selected stars in NGC\,6397.  
  The panels show the $H_{\rm E}$ vs.\,$I_{\rm E}-J_{\rm E}$ (left) and $H_{\rm E}$ vs.\,$Y_{\rm E}-H_{\rm E}$ (right) diagram for cluster members. The photometry, derived from Euclid data \citep{libralato2024a}, has been corrected for the differential-reddening effects.}  
  \label{fig:cmds}
\end{figure*} 

\section{Multiple populations in NGC 6397 from Euclid, HST and JWST}\label{sec:ngc6397}
To investigate multiple stellar populations in NGC\,6397 with Euclid data, we started analyzing the $H_{\rm E}$ vs.\,$I_{\rm E}-J_{\rm E}$ (left) and $H_{\rm E}$ vs.\,$Y_{\rm E}-H_{\rm E}$ CMDs plotted in Figure\,\ref{fig:cmds}. 
\begin{figure*} 
  \centering
 \includegraphics[height=9.5cm,trim={0.5cm 5.0cm 7.9cm 6.5cm},clip]{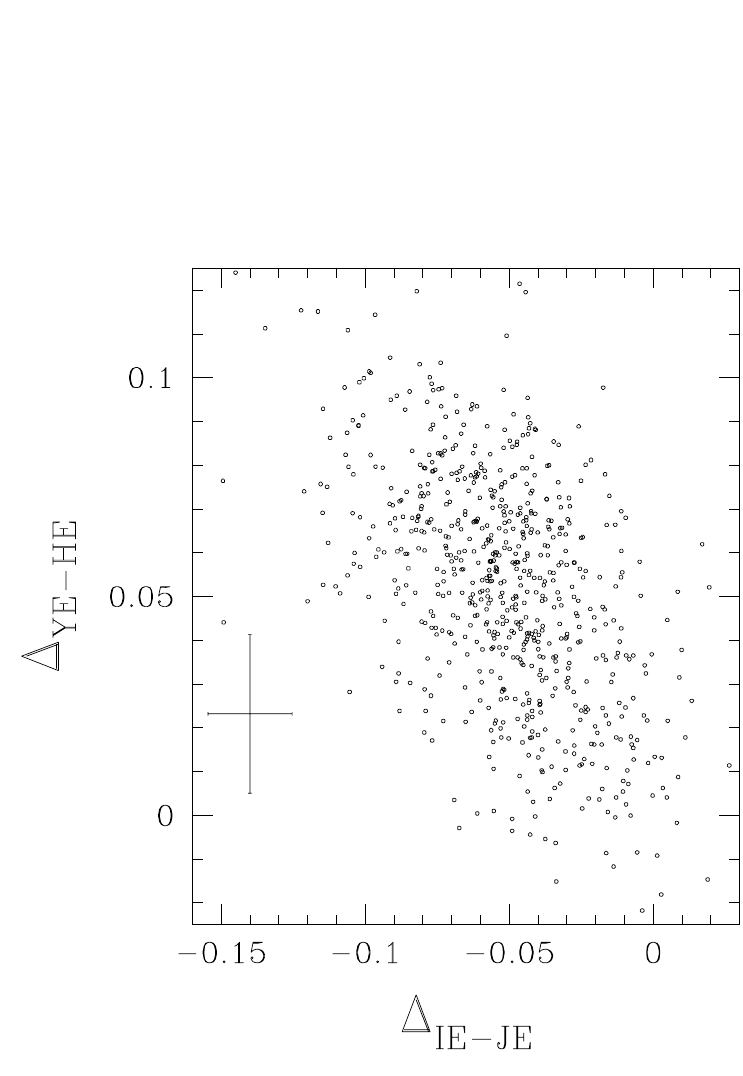}
  \includegraphics[height=9.5cm,trim={0.5cm 5.0cm 3.0cm 6.5cm},clip]{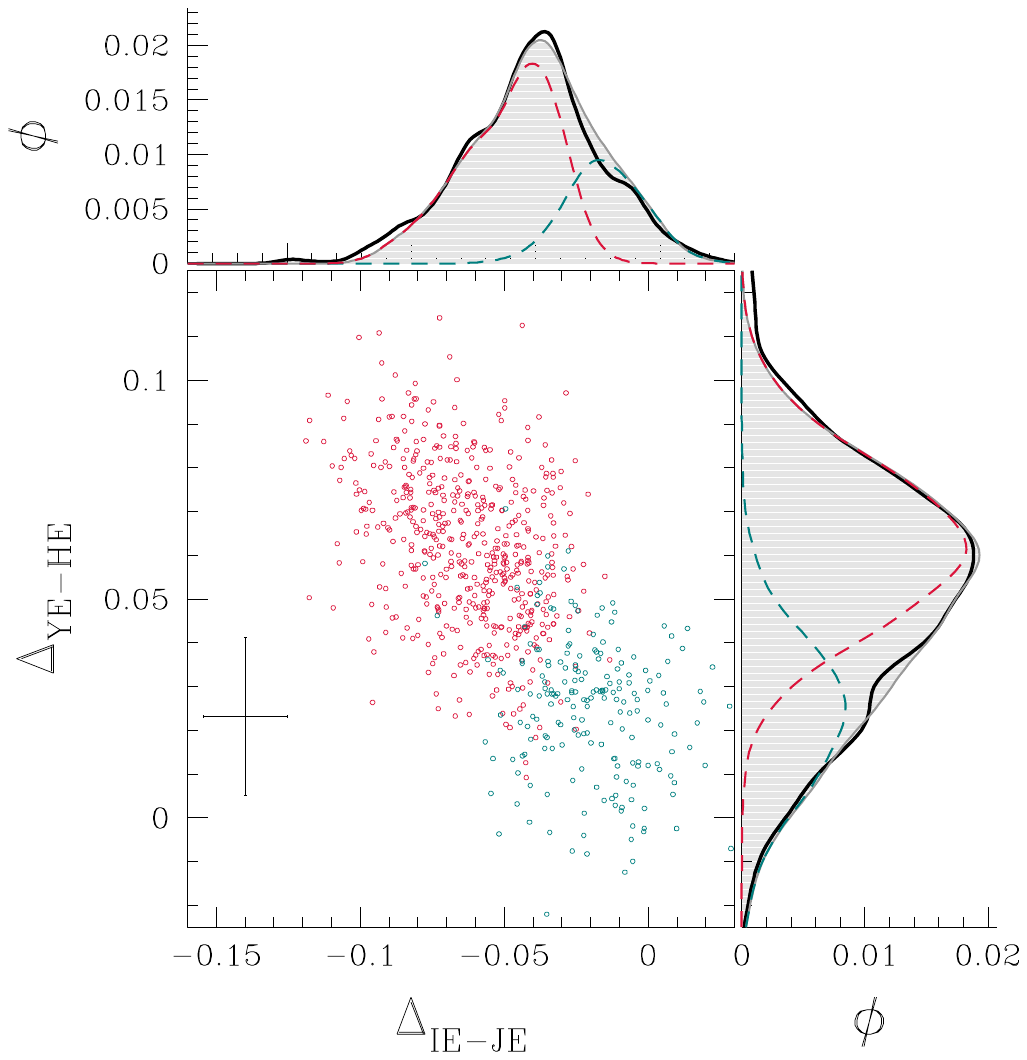}
  \caption{The left panel presents the observed ChM of NGC\,6397 derived from Euclid photometry, while the right panel shows the corresponding simulated ChM. The right panel also includes the kernel-density distributions of $\Delta_{\rm IE-JE}$ and $\Delta_{\rm YE-HE}$ for both the observed ChM (black lines) and the simulated ChM (gray lines with a shaded area). The red and blue dashed lines represent the simulated 1P and 2P stars, respectively}  
  \label{fig:chmD}
\end{figure*} 

The left panel of Figure\,\ref{fig:chmD} presents the $\Delta_{\rm IE-JE}$ vs.\,$\Delta_{\rm YE-HE}$ ChM of M dwarfs  derived from these CMDs within the magnitude range $19.8 < H_{\rm E} < 21.0$, where the impact of multiple populations on the MS color is most pronounced. This ChM, constructed following the methodology outlined by \citep[][see also Section\ref{sec:theory}]{milone2017b}, reveals two distinct stellar groups. Stars clustered near the origin of the reference frame correspond to the 1P, while the group located around ($\Delta_{\rm IE-JE}$, $\Delta_{\rm YE-HE}$) $\sim (-0.06, 0.06)$ represents the 2P.

The right panel of Figure\,\ref{fig:chmD} displays the simulated ChM that provides the best match with the observations and is derived as in \citet{zennaro2019a}. The 1P and 2P stars comprise the 31\% and 69\%, respectively, of the simulated stars and are colored teal and crimson, respectively. We also compare the kernel-density distributions of $\Delta_{\rm IE-JE}$ and $\Delta_{\rm YE-HE}$ for the observed (black line) and simulated stars (gray line) and indicate the distributions of 1P and 2P simulated stars with teal and crimson dashed lines, respectively.

\begin{figure*} 
  \centering
 \includegraphics[height=9.1cm,trim={1.2cm 9.5cm 0.0cm 4.5cm},clip]{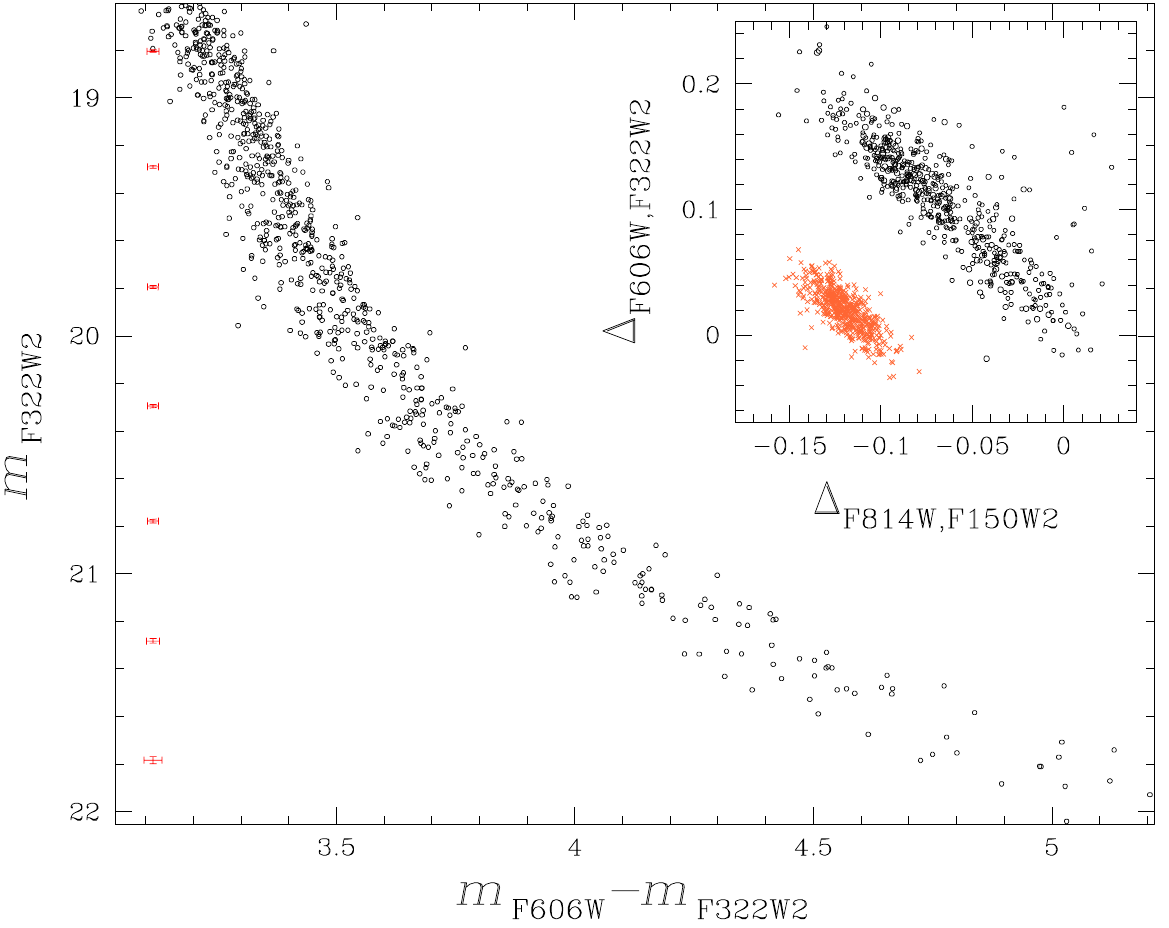}
  \caption{$m_{\rm F322W2}$ vs.\,$m_{\rm F606W}-m_{\rm F322W2}$ CMD of proper-motion-selected members of NGC\,6397, constructed using NIRCam data. The inset displays the $\Delta_{\rm F606W-F322W2}$ vs.\,$\Delta_{\rm F814W-F150W2}$ ChM, where black points represent NGC\,6397 stars, and the orange points correspond to a simulated ChM for a single-population cluster. Red error bars indicate the typical color and magnitude uncertainties for stars in different magnitude bins.}  
  \label{fig:F606WF322W2}
\end{figure*} 

To further investigate the multiple populations in NGC\,6397, we compare results obtained from JWST and Euclid data. Figure\,\ref{fig:F606WF322W2} presents the $m_{\rm F322W2}$ vs.\,$m_{\rm F606W}-m_{\rm F322W2}$ CMD, derived from NIRCam data and corrected for differential reddening. This CMD, along with the $m_{\rm F322W2}$ vs.\,$m_{\rm F814W}-m_{\rm F150W2}$ CMD, 
 has been utilized to construct the ChM shown in the inset.
A visual inspection of the ChM distinctly identifies two stellar populations: 1P stars, concentrated near the origin of the reference frame, and 2P stars, which exhibit significantly larger absolute values of $\Delta_{\rm F606W,F322W2}$ and $\Delta_{\rm F814W,F150W2}$. These results are in agreement with recent findings by \cite{scalco2024a}, who identified 1P and 2P stars among M-dwarfs of NGC\,6397 by using the same dataset.

\begin{figure*} 
  \centering
 \includegraphics[height=5.35cm,trim={0.5cm 5.35cm 0.1cm 12.5cm},clip]{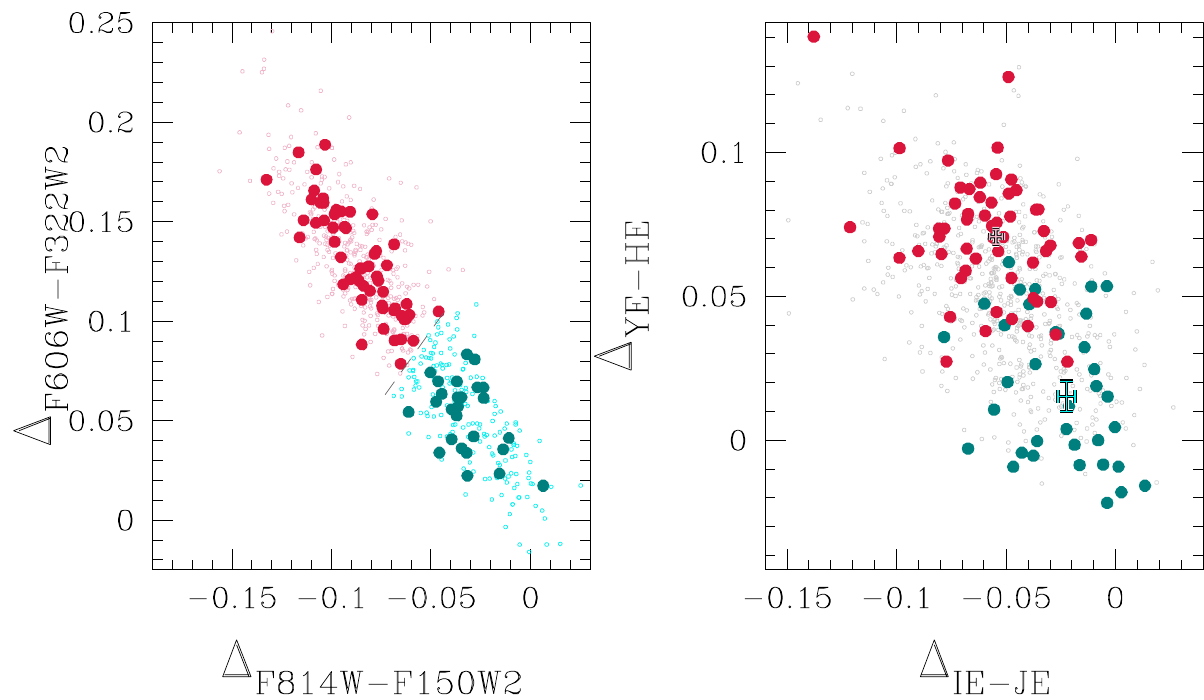}
  \includegraphics[height=5.35cm,trim={0.8cm 5.35cm 0.2cm 12.5cm},clip]{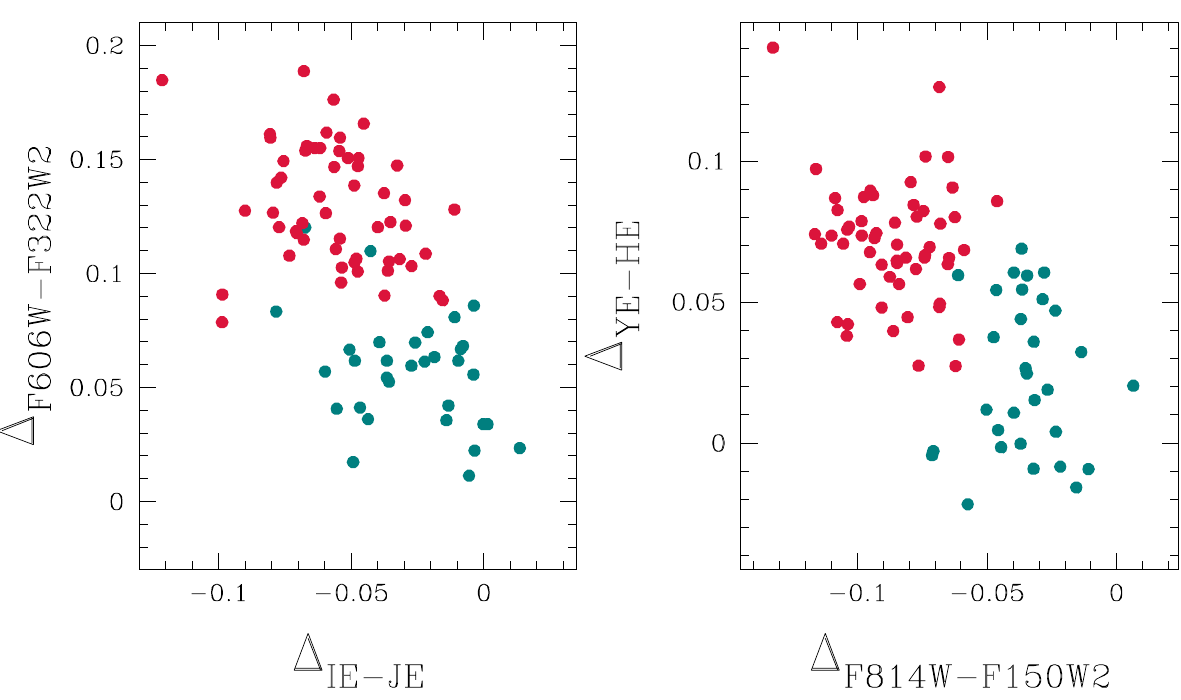}
 \caption{Collection of ChMs for M dwarfs in NGC\,6397. The leftmost panel reproduces the $\Delta_{\rm F606W-F322W2}$ vs.\,$\Delta_{\rm F814W-F150W2}$ ChM derived from NIRCam photometry, as presented in Figure\,\ref{fig:F606WF322W2}. In this panel, probable 1P and 2P stars are colored teal and crimson, respectively. The second panel from the left displays the ChM constructed from Euclid photometry, introduced in Figure\,\ref{fig:chmD}. Here, teal and crimson points with error bars indicate the average positions of 1P and 2P stars, respectively. The remaining panels present ChMs obtained by combining Euclid and NIRCam photometry. Large dots highlight stars with data available from both NIRCam and Euclid. }  
  \label{fig:mix}
\end{figure*} 

We used the ChM derived from NIRCam photometry to identify bona fide groups of 1P and 2P stars, represented by teal and crimson points, respectively, in the leftmost panel of Figure\,\ref{fig:mix}. Subsequently, we analyzed the distribution of these stars in other ChMs constructed using both NIRCam and Euclid photometry. The second panel from the left reproduces the $\Delta_{\rm IE-JE}$ vs.\,$\Delta_{\rm YE-HE}$ ChM introduced in Figure\,\ref{fig:chmD}, where teal and crimson dots represent the 1P and 2P stars, respectively, for which both Euclid and NIRCam photometry is available.
The observation that the majority of 1P and 2P stars, selected from the NIRCam ChM, occupy distinct regions in the Euclid ChM further strengthens the evidence for the presence of multiple populations in NGC\,6397. 
If the pseudo-color broadening in the ChMs were solely due to observational errors, the teal and crimson groups of stars, selected from NIRCam photometry, would have an equal likelihood of displaying large or small $\Delta_{\rm IE-JE}$ and $\Delta_{\rm YE-HE}$ values. However, the observation that these two groups exhibit distinct average pseudo-colors in the Euclid ChM, as indicated by the small points with error bars, provides strong evidence for intrinsic pseudo-color variations associated with multiple populations \citep{anderson2009a, milone2010a}.
For completeness, the two rightmost panels of Figure\,\ref{fig:mix} present two ChMs: $\Delta_{\rm F606W-F322W2}$ vs.\,$\Delta_{\rm IE-JE}$ and $\Delta_{\rm YE-HE}$ vs.\,$\Delta_{\rm F814W-F150W2}$, constructed using NIRCam and Euclid photometry. In both diagrams, the stars display an anticorrelation, with Spearman's rank correlation coefficients of approximately 0.6. This finding further corroborates the presence of 1P and 2P stars among M-dwarfs.

\begin{figure*} 
  \centering
 \includegraphics[height=9cm,trim={0.5cm 5.5cm 2.8cm 8.5cm},clip]{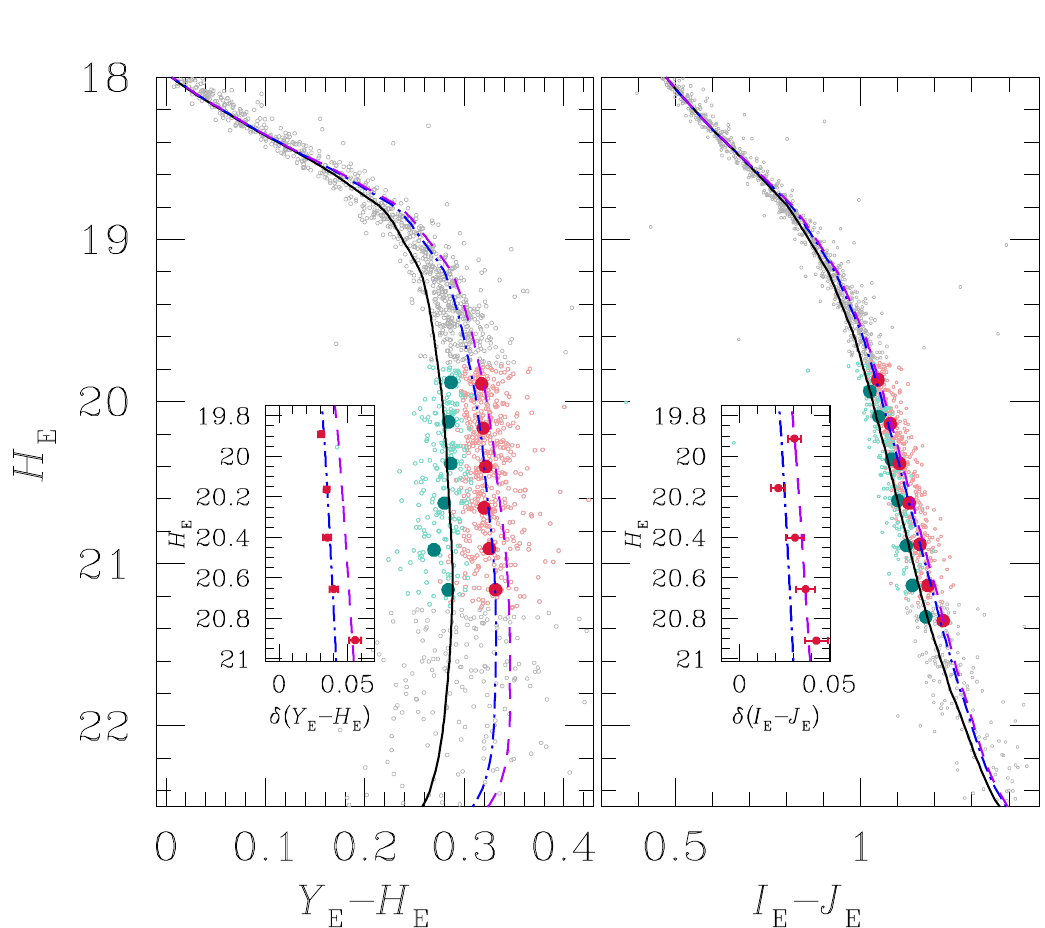}
 \includegraphics[height=9cm,trim={8.0cm 5.5cm 0.1cm 8.5cm},clip]{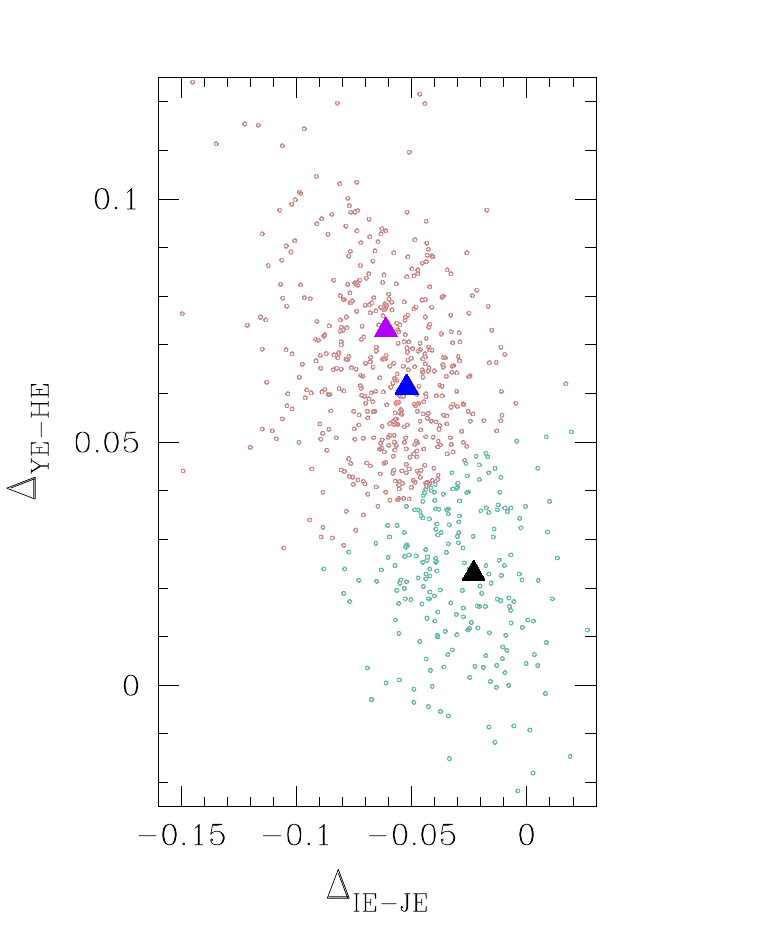}
 \caption{Reproductions of the CMDs (left and middle panels) and the ChM (right panel) for the proper-motion-selected NGC\,6397 cluster members from Figures\,\ref{fig:cmds} and \ref{fig:chmD} are presented. Stars selected from the ChM are color-coded, with light-green and light-red representing 1P and 2P stars, respectively. Isochrones from the BaSTI database \citep{pietrinferni2021a}, with the same iron abundance ([Fe/H]=$-$2.02) but varying oxygen contents of [O/Fe]=0.4, 0.1, and 0.0, are shown in black, blue, and purple, respectively. These isochrones represent 1P stars, and 2P stars with intermediate and large oxygen variations. Segments of the isochrones with $19.8<H_{\rm E}<21.5$ mag are highlighted as large triangles superimposed on the ChM.
Teal and crimson large dots overlaid on the CMDs represent the fiducial lines for 1P and 2P stars, respectively. The crimson dots in the insets of the left and middle panels indicate the color differences between the fiducials of 2P and 1P stars, $\delta$($Y_{\rm E}-H_{\rm E}$) and $\delta$($I_{\rm E}-E_{\rm E}$), at various $H_{\rm E}$ magnitude levels. The color differences as a function of magnitude for the O-intermediate and O-poor 2P isochrones are also displayed.}  
  \label{fig:ISOn6397}
\end{figure*} 

To infer the relative oxygen abundances of 1P and 2P stars using Euclid data alone, we followed the recipe by \cite{milone2023b}, leveraging the $H_{\rm E}$ vs.\,$Y_{\rm E}-H_{\rm E}$ and $H_{\rm E}$ vs.\,$I_{\rm E}-J_{\rm E}$ CMDs, corrected for differential reddening. These CMDs, reproduced in the left and middle panels of Figure\,\ref{fig:ISOn6397}, are compared with alpha-enhanced isochrones from the BaSTI database \citep{pietrinferni2021a}. The black isochrones correspond to [Fe/H]=$-$2.02 and Y=0.246 and are consistent with those used by \citet{massari2024a}.
In addition, we derived the colors of blue and magenta isochrones, which share the same iron abundance as the black isochrones but are depleted in oxygen by 0.3 and 0.4 dex, respectively. Considering the negligible helium difference between 1P and 2P stars in NGC\,6397 \citep[$\Delta Y=0.006 \pm 0.009$;][]{milone2018a}, we assumed the same helium content for all isochrones.
We determined five fiducial points corresponding to probable 1P and 2P stars, selected from the ChM in the right panel of Figure\,\ref{fig:ISOn6397}. These fiducials, spanning the $H_{\rm E}$ interval between 19.8 and 21.5 mag, are overlaid on the CMDs in this figure.

Although the black isochrones provide a reasonable match to the observed CMDs, small discrepancies between the fiducial points of 1P stars and the black isochrone are evident. To mitigate the impact of these discrepancies on the estimation of the relative abundances of multiple populations in NGC\,6397, we plotted in the insets of Figure\,\ref{fig:ISOn6397} the $Y_{\rm E}-H_{\rm E}$ and $I_{\rm E}-J_{\rm E}$ color differences between the 2P and 1P fiducials at various $H_{\rm E}$ magnitudes.
The blue and magenta lines in the insets represent the corresponding color differences between the O-poor isochrones and the black isochrone, showing a good match with the observed points. These isochrones are also superimposed on the ChM shown in the right panel of Figure\,\ref{fig:ISOn6397}, where they cluster into three distinct small regions marked with triangles.
Our analysis indicates that isochrones with [O/Fe]=0.1 provide the best match for the bulk of 2P stars, while isochrones with [O/Fe]=0.0 likely represent the 2P stars occupying the extreme positions in the ChM.

\subsection{Internal kinematics of stellar populations in NGC 6397}

The high-precision proper motions derived from Euclid, {\it HST}, and {\it JWST} data provide a unique opportunity to constrain the internal kinematics of NGC\,6397 and better understand various unresolved questions about its stellar populations. To demonstrate the potential of this dataset, we will analyze the proper motions of stars in the central field and in Fields A to E to investigate the internal kinematics of 1P and 2P stars, placing constraints on the degree of proper-motion anisotropy in each population. Additionally, leveraging the availability of high-precision proper motions for Field A stars —spanning a wide mass range from $\sim$0.1 to 0.8 solar masses— we aim to constrain the state of energy equipartition within this field.

\subsubsection{Proper motion anisotropy}\label{susbsub:anisotropy}
To study the internal dynamics of multiple populations in NGC\,6397, we followed the procedure outlined in various works on this topic \citep[e.g.][]{richer2013a,  cordoni2020a, cordoni2020b, libralato2022a, ziliotto2025a}. First, we converted the proper motions from Cartesian coordinates ($\mu_\mathrm{\alpha}\cos\delta$, $\mu_\mathrm{\delta}$) to radial and tangential components ($\mu_\mathrm{R}$, $\mu_\mathrm{T}$). Here, radial proper motions are defined as positive when moving outward, while tangential proper motions are positive for counter-clockwise rotation.

Next, we determined the velocity dispersions of 1P and 2P stars by considering the negative log-likelihood function: 
$$
\lambda=\prod_{i=1}^Np(v_i,\epsilon_i)
$$
 with the probability of finding a star with velocity $v_i$ and uncertainty $\epsilon_i$ given by:
$${}
p(v_i,\epsilon)=\frac{1}{2\pi\sqrt{(\sigma^2+\epsilon_i^2)}}\exp \left(-\frac{(v_i-v)^2}{2(\sigma^2+\epsilon^2_i)}\right)
$$
Finally, we found the intrinsic dispersion by maximizing the likelihood \citep[see][for details]{mackey2013a, marino2014a}.
 The uncertainties in the velocity dispersions were estimated using the Markov Chain Monte Carlo algorithm \texttt{emcee} \citep{foreman2013a}. Finally, we calculated the anisotropy parameter as $\beta = \sigma_\mathrm{T} / \sigma_\mathrm{R} - 1$.

Considering the varying number of stars in the studied fields, we analyzed the central field, field A, and field D individually, while fields B, C, and E have been combined to increase the statistical significance of the results. For fields with a sufficient number of stars, specifically, the central field and field A, we defined two radial bins, each containing an equal number of stars.

\begin{figure} 
  \centering
 \includegraphics[height=10cm,trim={0.0cm 0cm -1.cm 0.0cm},clip]{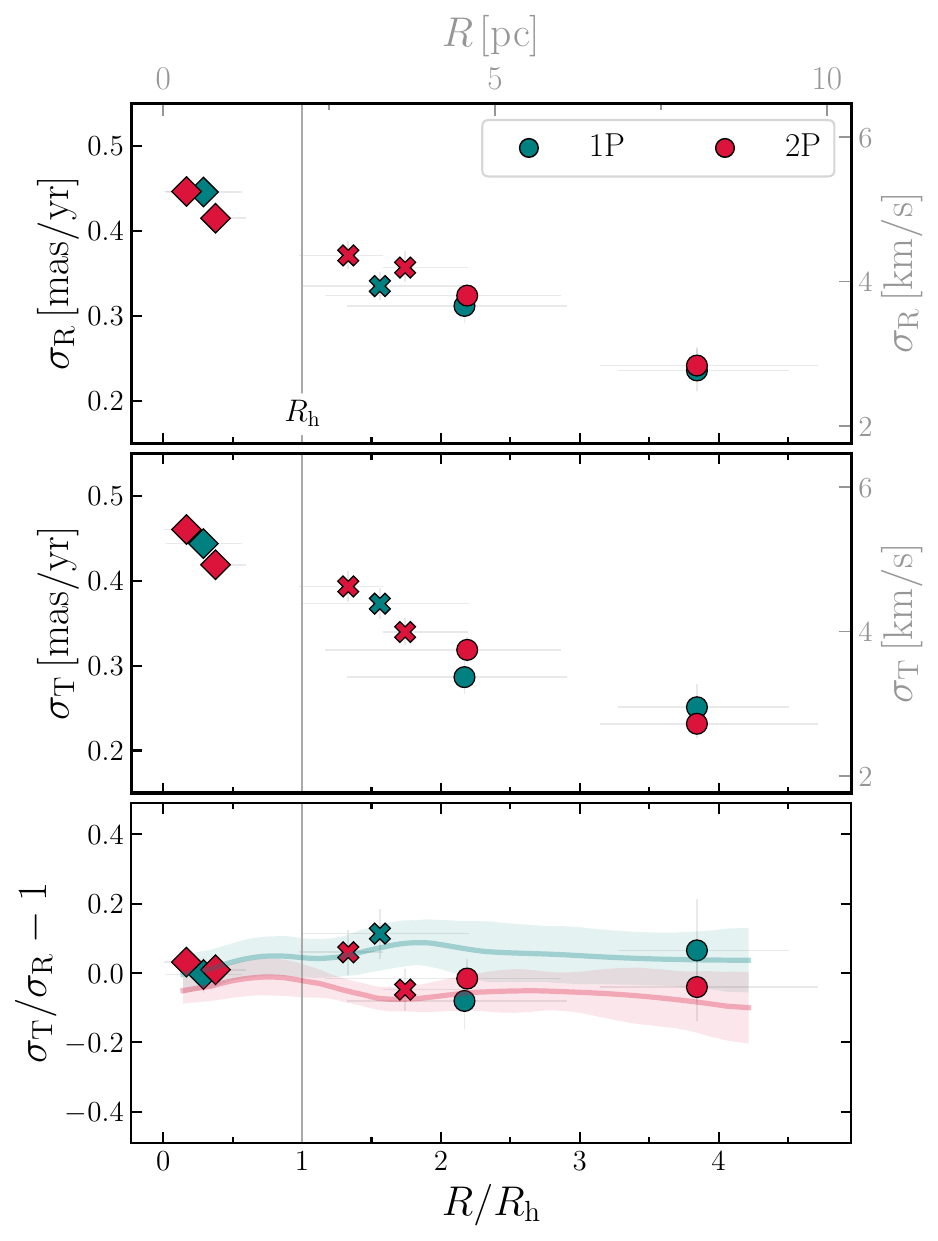}
 \caption{ Proper-motion dispersion along the radial (top panel) and tangential direction (middle panel) as a function of the distance from the  cluster center.  The lower panel shows the anisotropy  parameter against the radial distance.
 The vertical lines mark the half-light radius, whereas teal and crimson symbols represent 1P and 2P stars, respectively. The results obtained from the {\it HST}, Euclid, and {\it JWST} ChMs are plotted with diamonds, crosses, and dots, respectively.
  The distance from the cluster center is reported in units of half-light radius (bottom x-axis) and parsecs (top x-axis). Similarly, velocities are shown in mas/yr on the left y-axis and km/s on the right y-axis. Here, we assumed the cluster distance and half-light radius provided by the 2010 version of the \citet{harris1996a} catalog. The colored lines and shaded regions in the bottom panel represent the global anisotropy profiles of dynamically old clusters derived in \citet[][see their sections 3.2 and 4]{cordoni2024a}.}
 
  \label{fig:InternalDynamics}
\end{figure}

Our results for all fields are presented in Figure~\ref{fig:InternalDynamics}, with 1P stars shown in teal and 2P stars in crimson. The radial and tangential velocity dispersion profiles are displayed in the top and middle panels, respectively, while the anisotropy profile is shown in the bottom panel. 
We find that 1P and 2P stars exhibit similar radial and tangential velocity-dispersion profiles, with both populations displaying motion consistent with isotropy. These findings align with the average anisotropy distribution observed in dynamically old GCs, as inferred by \citet{cordoni2024a}. The corresponding distributions are represented by teal and crimson lines, along with shaded areas, in Figure\,\ref{fig:InternalDynamics}.

\subsubsection{Energy equipartition}
GCs may undergo a process known as two-body relaxation, which drives them toward energy equipartition. In this state, lower-mass stars attain higher velocity dispersions compared to their higher-mass counterparts \citep{spitzer1969a}. This phenomenon is typically quantified using the parameter $\eta$, which describes the relationship between velocity dispersion $\sigma_{\rm \mu}$ and stellar mass, M, as $\sigma_{\mu} \propto M^{\eta}$. Complete energy equipartition corresponds to $\eta$=0.5. However, N-body simulations reveal that GCs rarely achieve full equipartition, instead reaching a state of partial equipartition \citep{baumgardt2003a, trenti2013a}.

\begin{figure} 
  \centering
 \includegraphics[height=6cm,trim={1.0cm 5cm 0.cm 9.5cm},clip]{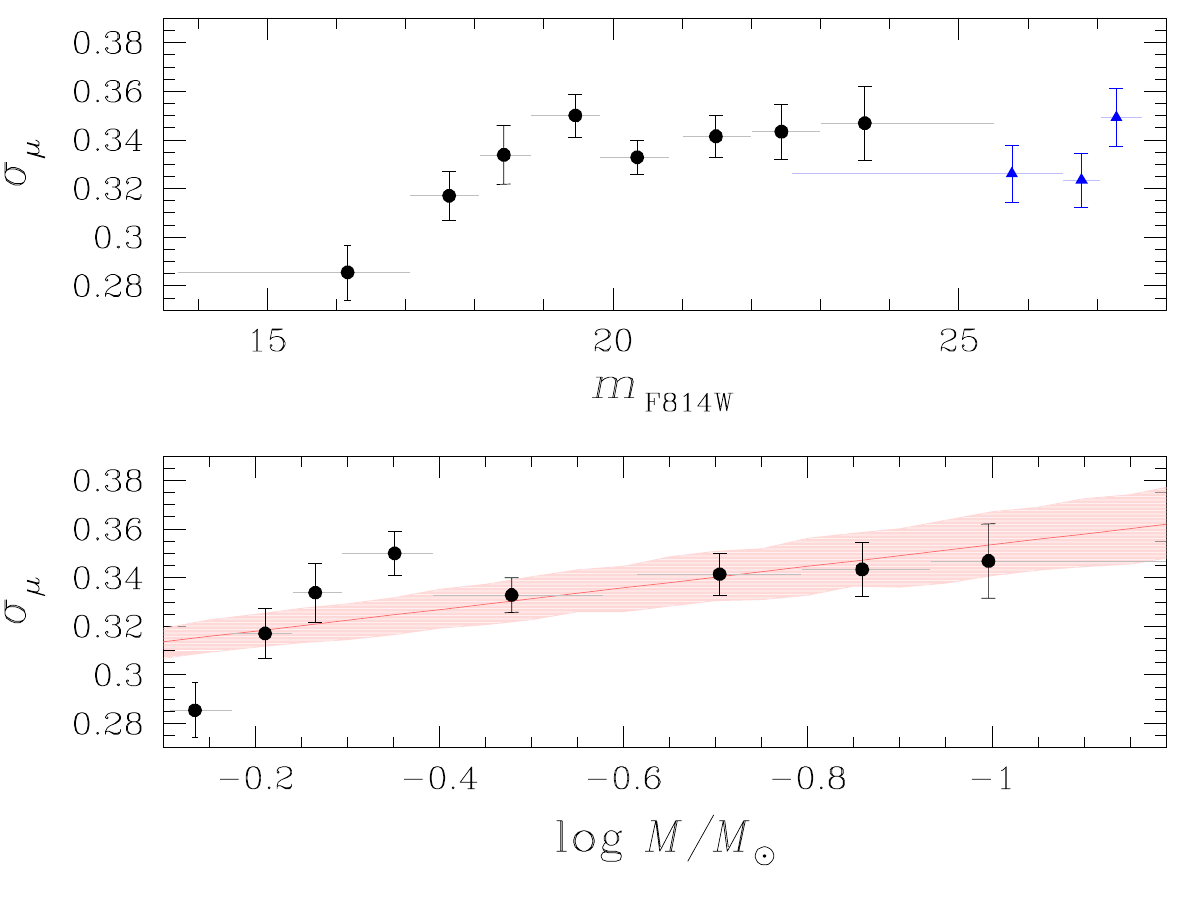}
 \caption{{\textit Top panel}. Velocity dispersion, $\sigma_{\rm \mu}$, as a function of the $m_{\rm F814W}$ magnitude for MS, SGB, and RGB stars (black circles) and for white dwarfs (blue circles). {\textit Bottom panel.} $\sigma_{\rm \mu}$ against the logarithm of stellar mass for MS, SGB, and RGB stars. We used light-red colors to represent the best-fit straight line. The horizontal segments indicate the magnitude and mass interval associated with each point. }  
  \label{fig:energy}
\end{figure}

To investigate the state of energy equipartion of stars with high-quality proper motion measurements in field A, we considered giants and MS stars in different bins of $m_{\rm F814W}$ and estimated the proper motion dispersion $\sigma_{\mu}$ as discussed in Section\,\ref{susbsub:anisotropy} and $\sigma_{\mu}$=$\sigma_{\rm T}=\sigma_{\rm R}$. We repeated the same analysis for white dwarfs.
The upper panel of Figure\,\ref{fig:energy} illustrates the proper motion dispersion  as a function of the F814W magnitude. A marginal increase of $\sim$0.05 mas yr$^{-1}$ is observed when comparing the velocity dispersion of giant stars to that of stars near the bottom of the MS. Additionally, white dwarfs exhibit a proper motion dispersion similar to that of most MS stars.
The bottom panel of Figure\,\ref{fig:energy} plots $\sigma_{\rm \mu}$  against the logarithm of stellar mass, which we inferred by using the best-fit isochrone from \citet{pietrinferni2021a}. The straight line represents the best-fit least-squares line, with a slope corresponding to $\eta=0.05\pm0.02$. This result indicates that the studied stars in NGC\,6397 exhibit only a very small degree of energy equipartition.

\subsection{The radial distribution of multiple populations}
To examine the radial distribution of 1P and 2P stars in NGC\,6397, we utilized ChMs constructed from Euclid, {\it HST}, and {\it JWST} photometry. The $\Delta_{\rm YE-HE}$ vs.\,$\Delta_{\rm IE-JE}$ Euclid ChM was derived using M-dwarfs identified as probable cluster members based on the positional displacements between the catalogs of \citet{stetson2019a} and \citet{libralato2024a}. These stars are marked as black points in the middle and right panels of Figure\,\ref{fig:ngc6397_rmap}.

\begin{figure} 
  \centering 
 \includegraphics[height=4.75cm,trim={0.8cm 5.2cm 6.cm 10.5cm},clip]{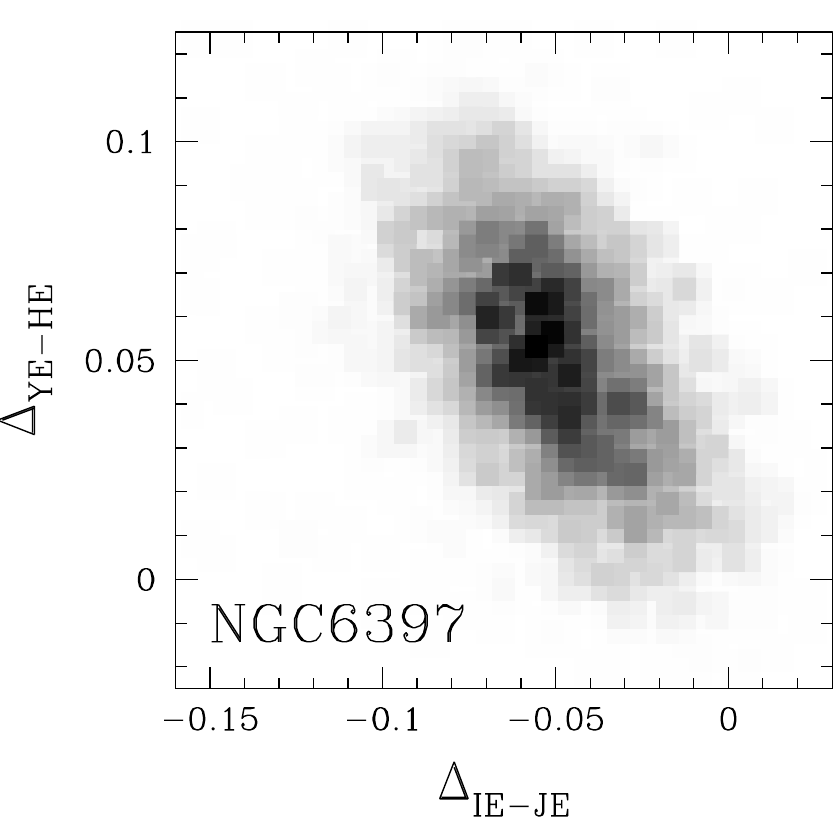}
 \includegraphics[height=4.75cm,trim={3.7cm 5.2cm 6.cm 10.5cm},clip]{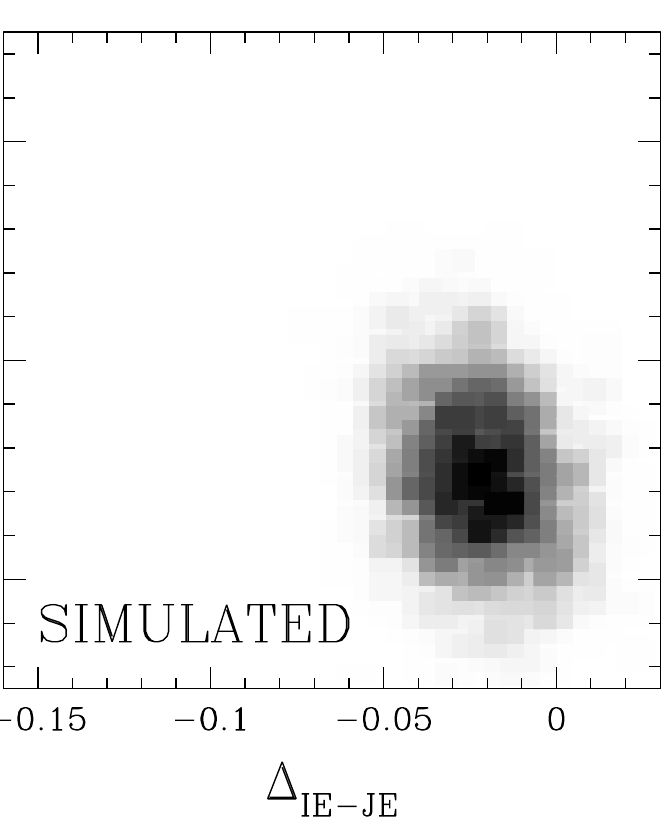}
 \caption{The left panel presents the Hess diagram of the $\Delta_{\rm YE-HE}$ vs.\,$\Delta_{\rm IE-JE}$ ChM constructed from Euclid photometry for NGC\,6397 stars. The stars were selected based on positional displacements between the catalogs of \citet{stetson2019a} and \citet{libralato2024a}. For comparison, the right panel shows the same ChM but exclusively for simulated 1P stars, highlighting the expected distribution of a single population. } 
  \label{fig:ChMall}
\end{figure}

The left panel of Figure\,\ref{fig:ChMall} displays the Hess diagram of this ChM, which includes stars spanning a broad field of view, from less than 2 arcminutes to 28 arcminutes from the center of NGC\,6397.

For comparison, the right panel of Figure\,\ref{fig:ChMall} shows the Hess diagram for simulated 1P stars only, constructed by incorporating observational photometric uncertainties. This simulated diagram represents the expected distribution of stars in a single-population cluster, underscoring the presence of multiple populations in NGC\,6397.

\begin{figure} 
  \centering
 \includegraphics[height=10cm,trim={0.8cm 5.2cm 6.5cm 5.0cm},clip]{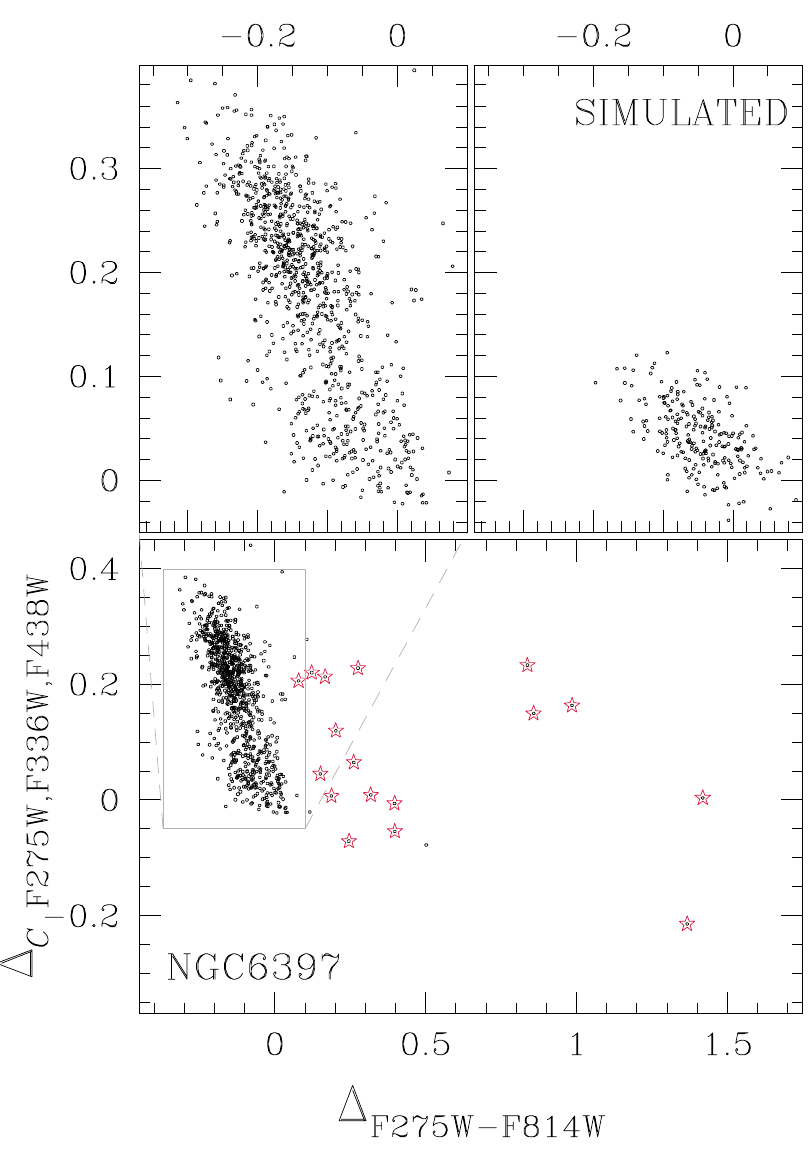}
 \caption{ChM of NGC\,6397 MS stars in the central field from {\it HST} photometry.
The ChM shown in the bottom panel includes the MS binaries (starred symbols), whereas the top-left panel is a zoom on the ChM region occupied by single stars.
 Top-right panel shows the simulated ChM for 1P stars.} 
  \label{fig:ChMcenter}
\end{figure} 
To explore the central region of NGC\,6397, we derived the $\Delta_{C,{\rm F275W,F336W,F438W}}$ vs.\,$\Delta_{\rm F275W-F814W}$ ChM for MS stars in the central field, as illustrated in Figure\,\ref{fig:ChMcenter}. The bottom panel of this figure includes all MS stars, with probable binaries having large mass ratios (q$\gtrsim$0.5) highlighted by crimson star symbols. These binaries were identified based on their positions in the $m_{\rm F814W}$ vs.\,$m_{\rm F606W}-m_{\rm F814W}$ and $m_{\rm F814W}$ vs.\,$m_{\rm F438W}-m_{\rm F814W}$ CMDs.

The top panels compare the observed ChM for probable single stars with the simulated ChM for 1P stars only, emphasizing the predominance of 2P stars among the analyzed stars in NGC\,6397.

While a quantitative assessment of the binary fraction among 1P and 2P stars is outside the scope of this study, the observation that binaries with similar $\Delta_{\rm F275W-F814W}$ values span a wide range of $\Delta_{C,{\rm F275W,F336W,F438W}}$ suggests that the cluster core hosts comparable fractions of binary systems composed of both 1P and 2P stars.
\begin{figure} 
  \centering
 \includegraphics[height=7.5cm,trim={1.0cm 5.6cm 1.3cm 8.5cm},clip]{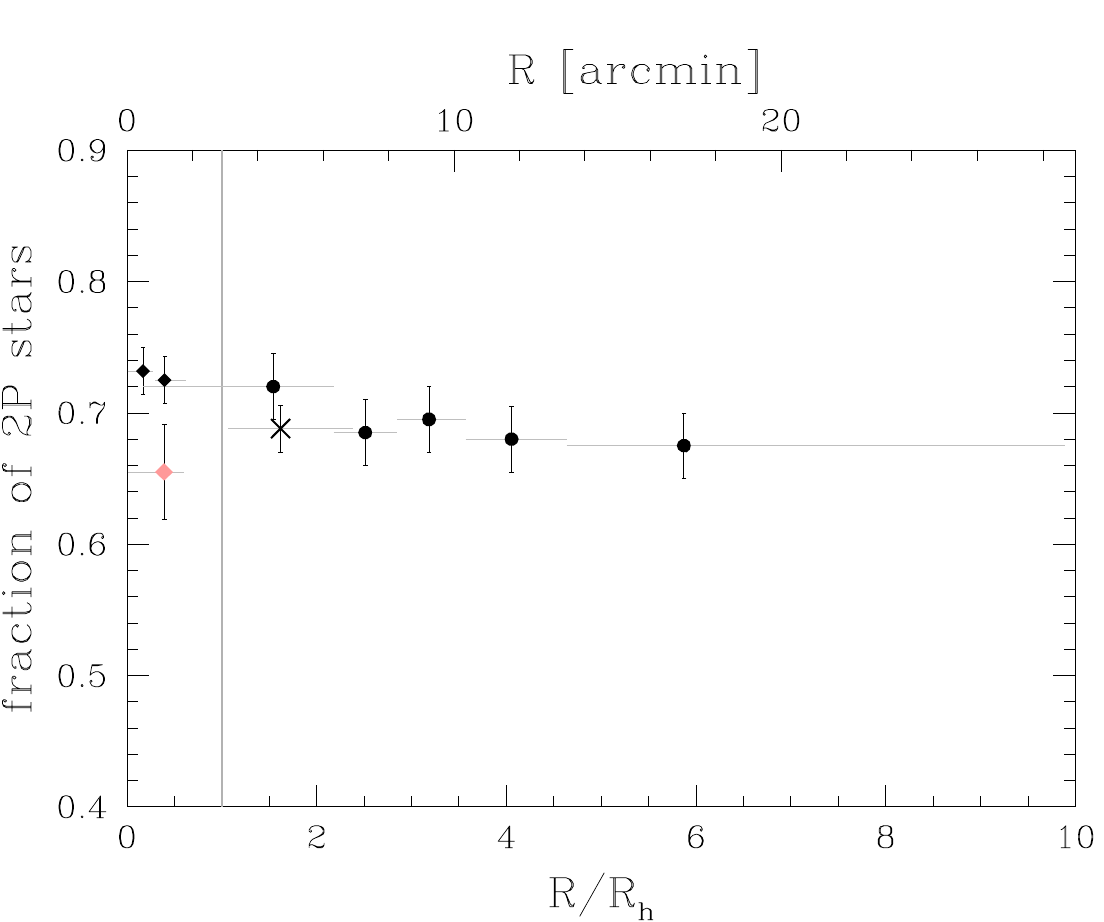}
 \caption{Fraction of 2P stars as a function of radial distance from the cluster center. The radial scale is provided both in units of the half-light radius \citep[$R_{\rm h}=2.9$ arcmin; 2010 version of the][catalog]{harris1996a} and in arcminutes. Diamonds, crosses, and large dots represent results derived from {\it HST}, combined {\it JWST} and {\it HST}, and Euclid data, respectively. Results from this study are shown in black, while the fraction of 2P stars determined by \citet{milone2017a} for RGB stars is depicted with a light-red diamond. Horizontal lines indicate the radial range associated with each determination of the 2P fraction.} 
  \label{fig:RD}
\end{figure} 
Furthermore, we used the $\Delta_{\rm F606W-F322W2}$ vs.\,$\Delta_{\rm F814W-F150W2}$ ChM derived from NIRCam and ACS/WFC photometry of stars in field A, as introduced in Figure\,\ref{fig:F606WF322W2}.

We utilized each ChM to estimate the fractions of 1P and 2P stars by following the procedure outlined by \citet{zennaro2019a}, which was also applied in Section\,\ref{sec:ngc6397}. To better examine the population ratio as a function of radial distance, we defined five radial bins containing equal numbers of M-dwarfs used for the Euclid ChM, and two additional radial bins with equal numbers of bright MS stars used for the central-field ChM. Each bin was analyzed separately.

The resulting fractions of 2P stars are shown with black symbols in Figure\,\ref{fig:RD}, alongside the 2P fraction derived by \citet{milone2017a} for RGB stars, represented by a light-red symbol. Notably, the fraction of 2P stars inferred from MS stars in the central field and field A agrees, within one sigma, with the results by \citet{milone2012a} and \citet{scalco2024a} obtained from similar datasets. However, in the central field, the bright-MS star sample yields larger 2P fractions compared to the RGB results by \citet{milone2017a}, with agreement at only the $\sim$1.5-sigma level.

Overall, we found that the fraction of 2P stars is approximately $\sim$70\% and remains nearly constant across the analyzed radial range. There is a slight indication that 2P stars may be slightly more concentrated, with their fraction decreasing by less than $\sim$4\% from the cluster center to the outer regions.

\section{Summary and conclusions}\label{sec:conclusions}
We present the first study investigating multiple stellar populations in GCs using Euclid data.
Our analysis begins with a theoretical exploration of how stellar populations with varying chemical compositions affect stellar magnitudes in the Euclid bands. To this end, we generated synthetic spectra for MS, SGB, and RGB stars with different helium, carbon, nitrogen, and oxygen contents. We examined metal-poor, metal-intermediate, and metal-rich clusters with [Fe/H]=$-2.30$, $-1.50$, and $-0.75$, respectively.

The fluxes of bright MS stars and giants are only minimally influenced by variations in C, N, and O abundances. In contrast, for M-dwarfs, the most pronounced differences between the spectra of 2P-like stars with enhanced C, N, and O and those of 1P-like stars with similar atmospheric parameters arises at wavelengths $\lambda \gtrsim 13,000$\AA\,. These differences, driven by molecular absorption involving oxygen, primarily affect the $H_{\rm E}$ band and, to a lesser extent, the $J_{\rm E}$ band.
In addition, in the metal-rich populations we observe strong molecular absorption features affecting the optical portion of the M-dwarf spectra, which are responsible for large $I_{\rm E}$ magnitude differences between 2P and 1P stars.

We used the synthetic spectra to derive the colors and magnitudes of Dartmouth isochrones that mimic 1P and 2P stars. For [Fe/H]=$-$1.50 and $-$2.30, the CMDs constructed with the $Y_{\rm E}-H_{\rm E}$ color provide the optimal separation between the M-dwarf sequences of 1P and 2P stars for all adopted helium abundances. The $I_{\rm E}-J_{\rm E}$ color serves as an alternative for disentangling multiple populations among M-dwarfs, albeit in stellar populations with large helium contents, the helium and oxygen variations have opposite effects on stellar colors.
As a consequence, the $\Delta_{\rm YE}-\Delta_{\rm HE}$ vs.\,$\Delta_{\rm IE}-\Delta_{\rm JE}$ would be an efficient tool to identify 1P and 2P stars among M-dwarfs.

In the case of metal-rich isochrones, due to the rise of molecular absorptions at wavelengths shorter than $\sim$7,000 \AA, the $I_{\rm E}$ band becomes very sensitive to the oxygen abundance of the M-dwarfs. At odds with what is observed at lower metallicities, the CMDs constructed with this filter are very efficient tools to identify multiple stellar populations in GCs. This phenomenon is similar to what is observed with {\it HST} filters, where CMDs constructed with the F606W band, whose transmission curve partially overlaps with the Euclid $I_{\rm E}$ band, such as $m_{\rm F814W}$ vs.\,$m_{\rm F606W}-m_{\rm F814W}$, show nearly coincident M-dwarf sequences in metal-poor GCs, such as NGC\,6397 (see e.g.\,Figure\,\ref{fig:cmdJWSTHST}), but exhibit large color separations in metal-rich GCs such as 47\,Tucanae \citep[e.g.][]{richer2013a, milone2022a, marino2024a}. Moreover, the $Y_{\rm E}-H_{\rm E}$ color is an efficient tool to disentangle 1P and 2P stars.

As a consequence, various photometric diagrams such as the $\Delta_{\rm YE-HE}$ vs.\,$\Delta_{\rm IE-YE}$ or $\Delta_{\rm YE-HE}$ vs.\,$\Delta_{\rm IE-JE}$ ChM and the pseudo-CMDs constructed with the $C_{\rm IE,YE,HE}$=($I_{\rm E}-Y_{\rm E}$)-($Y_{\rm E}-H_{\rm E}$) index are optimal diagrams to disentangle multiple populations in metal-rich GCs.

While variations in C, N, and O do not significantly affect the colors of stars brighter than the MS knee, helium differences are responsible for the splitting of MS and faint RGB stars in CMDs constructed with the $I_{\rm E}$ band. Helium-rich stars are hotter than stars with 1P-like helium abundance; hence wide color baselines, such as $I_{\rm E}-H_{\rm E}$, would allow the detection of distinct populations with different helium content.

This effect of helium on stellar color has been widely used in studies on multiple populations. The first evidence of multiple MSs in GCs has been provided by CMDs constructed with optical colors such as the $m_{\rm F475W}-m_{\rm F814W}$ or $m_{\rm F555W}-m_{\rm F814W}$ colors from {\it HST} \citep[e.g.][]{anderson1997a,bedin2004a,dantona2005a,piotto2007a,milone2010a}. Recent surveys of multiple populations in GCs with {\it HST} and ground-based telescopes utilize the $\Delta_{C {\rm F275W,F336W,F438W}}$ vs.\,$\Delta_{\rm F275W-F814W}$ and $\Delta_{C {\rm U,B,I}}$ vs.\,$\Delta_{\rm B-I}$ ChM, respectively \citep{milone2017a, jang2022a}. The x-axis of these ChMs is derived from the $m_{\rm F275W}-m_{\rm F814W}$ and $B-I$ colors, and is very sensitive to helium variations in GCs. In this context, the Euclid NIR magnitudes, together with far-UV or blue filters, would provide wider color baselines than those commonly used in the literature. As a consequence, the $\Delta_{C {\rm F275W,F336W,F438W}}$ vs.\,$\Delta_{\rm F275W-HE}$ and $\Delta_{C {\rm U,B,I}}$ vs.\,$\Delta_{\rm B-HE}$ ChMs would improve our capability to detect multiple stellar populations along the upper MS and the RGB.

Inspired by our theoretical findings, we explore multiple populations in the nearby GC NGC\,6397, recently observed by Euclid \citep{massari2024a}. We combine stellar astrometry and photometry obtained from Euclid data \citep{libralato2024a} with observations from various facilities.

Specifically, we used archival {\it HST} and {\it JWST} images to derive high-precision photometry and astrometry of stars in six distinct fields. The stellar positions from these data, combined with those from Euclid, allowed us to derive accurate stellar proper motions, enabling the discrimination of cluster members from field stars and facilitating the investigation of the internal kinematics of NGC\,6397 stars.
Further selection of cluster members across a wide field of view was achieved by analyzing the displacements between stellar positions in the Euclid catalog and the astro-photometric catalog provided by Peter Stetson.

The main results on NGC\,6397 include:
\begin{itemize}
\item 
 The $\Delta_{\rm YE-HE}$ vs.\,$\Delta_{\rm IE-JE}$ ChM constructed from the $H_{\rm E}$ vs.\,$I_{\rm E}-J_{\rm E}$ and $H_{\rm E}$ vs.\,$Y_{\rm E}-H_{\rm E}$ CMDs of proper-motion selected cluster members exhibit significant color broadening among M-dwarfs, which is consistent with the existence of two stellar populations. Approximately 30\% of the studied M-dwarfs belong to the 1P population, with pseudo-color broadening consistent with observational errors. In contrast, the 2P stars display an intrinsic color broadening, indicative of small but significant star-to-star chemical inhomogeneities.

 \item For stars in field A, we obtained deep photometry in the F606W and F814W bands of WFC/ACS and the F150W2 and F322W2 filters of NIRCam. Consistent with previous studies \citep[][]{richer2006a, bedin2024a, scalco2024a, gerasimov2024a}, these data enabled us to construct deep CMDs of NGC\,6397 stars, revealing the faintest cluster members. For instance, the $m_{\rm F814W}$ vs.\,$m_{\rm F814W}-m_{\rm F150W2}$ CMD reaches the end of the white dwarf cooling sequence, showing sharp changes of the sequence slope at $m_{\rm F814W} \sim 26.7$ and 27.5 mag. Similarly, the $m_{\rm F322W2}$ vs.\,$m_{\rm F150W2}-m_{\rm F322W2}$ CMD approaches the hydrogen-burning limit of MS stars and reveals the upper portion of the brown-dwarf sequence.
Despite the small number of faint MS stars, these diagrams hint at potential gaps among very low-mass stars, as recently observed in 47\,Tucanae by \citet{marino2024a}.

\item {\it HST} and {\it JWST} photometry of stars in field A is used to derive the $\Delta_{\rm F606W-F322W2}$ vs.\,$\Delta_{\rm F814W-F150W2}$ ChM, which clearly reveals the presence of 1P and 2P stars. The clustering of 1P and 2P stars, as identified in the ChM constructed from NIRCam and ACS/WFC photometry, into distinct regions of the Euclid ChM further highlights the capability of Euclid photometry to disentangle multiple stellar populations in GCs.

\item We compared the distribution of 1P and 2P stars in the Euclid CMDs ($H_{\rm E}$ vs.\,$I_{\rm E}-J_{\rm E}$ and $H_{\rm E}$ vs.\,$Y_{\rm E}-H_{\rm E}$) and in the $\Delta_{\rm YE}-\Delta_{\rm HE}$ vs.\,$\Delta_{\rm IE-\rm JE}$ ChM with $\alpha$-enhanced isochrones from the BaSTi database \citep{pietrinferni2021a}, which account for the chemical composition of multiple populations in NGC\,6397. The observations indicate that the bulk of 2P stars are depleted by $\sim$0.3 dex in [O/Fe] compared to the 1P stars, which exhibit [O/Fe]=0.4. Additionally, a small fraction of 2P stars, comprising about one-fifth of the total 2P population, show lower oxygen abundances with [O/Fe]$\sim$0.0. 

The average oxygen difference between 1P and 2P stars is consistent with the findings of \citet{carretta2005a}, based on spectroscopy of RGB stars, and \citet{scalco2024a}, who used NIRCam photometry of M-dwarfs, thus demonstrating that Euclid photometry is a valuable tool to estimate the content of light elements in 2P stars relative to the 1P. However, \citet{lind2011a} derived a smaller [O/Fe] difference between 1P and 2P stars of $\sim$0.2 dex by using high-precision spectroscopy of RGB stars. The discrepancy between our results on the oxygen abundances of multiple populations in NGC\,6397 and those by Lind and collaborators may be partially reconciled by considering that the oxygen content of 1P stars assumed in our study is approximately 0.3 dex lower than the value inferred by \citet{lind2011a}.

\item The wide field of view of Euclid has enabled an extensive investigation into the radial distribution of multiple populations in NGC\,6397. Using Euclid photometry, we identified multiple populations among M-dwarfs spanning a radial range from $\sim$2 to $\sim$28 arcmin from the cluster center. In the crowded central regions, where disentangling 1P and 2P stars with Euclid data is challenging, we utilized the ChM of bright MS stars constructed from F275W, F336W, F438W, and F814W images obtained with the UVIS/WFC3 and WFC/ACS cameras on {\it HST}. Our analysis reveals that the fraction of 1P stars remains nearly constant across the studied radial interval. However, there is a subtle indication of a possible gradient, with the fraction of 2P stars increasing by less than $\sim$4\% when moving from the outermost cluster regions toward the center.

   \item Finally, we derived high-precision stellar proper motions by comparing stellar positions in Euclid images with archival images from {\it HST} and {\it JWST}. This enabled us to investigate the radial and tangential proper-motion dispersions of multiple populations at various distances from the cluster center. Our analysis reveals that 1P and 2P stars exhibit similar proper-motion distributions. Both populations display isotropic proper motions, consistent with the behavior typically observed in dynamically evolved GCs \citep{cordoni2024a}.
 
\end{itemize}

In summary, our theoretical analysis of multiple populations using Euclid data has identified the CMDs and ChMs best suited for distinguishing 1P and 2P stars. In particular, Euclid photometry proves to be an efficient tool for identifying multiple stellar populations in the low-mass regime of M dwarfs and inferring their oxygen abundances.

Unlike studies based on {\it HST} and {\it JWST}, Euclid’s wide field of view makes it ideally suited for analyzing multiple stellar populations across broad spatial extents and at varying radial distances from cluster centers. This capability provides crucial insights for constraining scenarios related to the formation and evolution of this elusive phenomenon.

These theoretical conclusions are strongly supported by our findings on NGC\,6397. For the first time, we have identified the multiple populations of M dwarfs over a wide field of $\sim$28 arcmin from the cluster center, constrained their radial distribution, and explored their internal kinematics.

\begin{acknowledgements}
This work has received funding from 
"PRIN 2022 2022MMEB9W - {\it Understanding the formation of globular clusters with their multiple stellar generations}" (PI Anna F.\,Marino), 
from INAF Research GTO-Grant Normal RSN2-1.05.12.05.10 -  (ref. Anna F. Marino) of the "Bando INAF per il Finanziamento della Ricerca Fondamentale 2022", and from the European Union’s Horizon 2020 research and innovation programme under the Marie Skłodowska-Curie Grant Agreement No. 101034319 and from the European Union – NextGenerationEU (beneficiary: T. Ziliotto).
\end{acknowledgements}

\section*{Data Availability}
The data underlying this article will be shared on reasonable request with the corresponding author.

\bibliographystyle{aa}
\bibliography{ms}

 \begin{appendix}
\section{HST and JWST images}
The Table\,\ref{tab:data} provides the main information on the images of NGC\,6397 collected by {\it HST} and {\it JWST} and used in the paper. In addition we provide the coordinates of each studied field.
\begin{table*}\label{tab:data}
\caption{Summary information about the {\it HST} and {\it JWST} images used in this work.}
    \centering
    \begin{tabular}{lccccccc}
    \\
 Date & N $\times$ exposure time & Mission & Instrument & Filter & Program & PI   \\
    \hline
                &               &           &         &        &      &      \\
                  & Central Field & RA=17$^h$40$^m$42$^s$  & Dec=$-$53$^{o}$39$^{'}$58$^{"}$  &       &              \\
2006 May 29  & 1s$+$4$\times$15s     & {\it HST} & WFC/ACS & F606W & 10775 & A.\,Sarajedini\\
2006 May 29  & 1s$+$4$\times$15s     & {\it HST} & WFC/ACS & F814W & 10775 & A.\,Sarajedini\\
2014 March 27  & 1504s     & {\it HST} & UVIS/WFC3 & F275W & 13297 & G.\,Piotto\\
2014 June 11  & 1418s     & {\it HST} & UVIS/WFC3 & F275W & 13297 & G.\,Piotto\\
2014 March 27  & 620s     & {\it HST} & UVIS/WFC3 & F336W & 13297 & G.\,Piotto\\
2014 June 11  & 620s     & {\it HST} & UVIS/WFC3 & F336W & 13297 & G.\,Piotto\\
2014 March 27  & 66s     & {\it HST} & UVIS/WFC3 & F438W & 13297 & G.\,Piotto\\
2014 June 11  & 66s     & {\it HST} & UVIS/WFC3 & F438W & 13297 & G.\,Piotto\\

 2016 March 03 & 4$\times$249s & {\it HST} & IR/WFC3 & F110W & 14124 & M.\,Correnti \\
 2016 March 03 & 4$\times$399s & {\it HST} & IR/WFC3 & F160W & 14124 & M.\,Correnti \\\\

                   & Field A & RA=17$^h$41$^m$03$^s$  & Dec=$-$53$^{o}$44$^{'}$38$^{"}$  &       &              \\
2005 March 13  & 2$\times$1s$+$2$\times$5s$+$2$\times$40s$+$       & {\it HST} & WFC/ACS & F606W & 10424 & R.\,H.\,Richer \\
- April 08 & 126$\times$616-769s & &  &  & &  \\
2005 March 13  & 2$\times$1s$+$2$\times$5s$+$2$\times$40s$+$       & {\it HST} & WFC/ACS & F814W & 10424 & R.\,H.\,Richer \\
 - April 08 & 252$\times$584-804s & &  &  & &  \\
 2010 March 09-11 & 3$\times$5s$+$3$\times$10s$+$3$\times$40s$+$       & {\it HST} & WFC/ACS & F814W & 11633 &  M.\,R.\,Rich\\
 & 6$\times$1280s$+$12$\times$1405s & &  &  & &  \\
 
 2014 March 27  & 683s     & {\it HST} & WFC/ACS & F475W & 13297 & G.\,Piotto\\
2014 March 27  & 610s     & {\it HST} & WFC/ACS & F814W & 13297 & G.\,Piotto\\\\

 2023 March 14 & 12$\times$623s       & {\it JWST} & NIRCam & F150W2 & 1979 &  L.\,Bedin\\
 2023 March 14 & 12$\times$623s       & {\it JWST} & NIRCam & F322W2 & 1979 &  L.\,Bedin\\\\

                   & Field B & RA=17$^h$41$^m$28$^s$  & Dec=$-$53$^{o}$49$^{'}$10$^{"}$  &       &              \\
2003 March 27-28  & 400s$+$558s$+$3$\times$700s      & {\it HST} & WFC/ACS & F775W & 9480 & J.\,D.\,Rhodes \\\\

                   & Field C & RA=17$^h$40$^m$45$^s$  & Dec=$-$53$^{o}$51$^{'}$32$^{"}$  &       &              \\
2003 May 16  & 327s$+$8$\times$400s      & {\it HST} & WFC/ACS & F775W & 9480 & J.,D.\,Rhodes \\\\

                   & Field D & RA=17$^h$40$^m$14$^s$  & Dec=$-$53$^{o}$44$^{'}$37$^{"}$  &       &              \\
2014 June 11  & 640s     & {\it HST} & WFC/ACS & F475W & 13297 & G.\,Piotto\\
2014 June 11  & 567s     & {\it HST} & WFC/ACS & F814W & 13297 & G.\,Piotto\\\\

                   & Field E & RA=17$^h$39$^m$23$^s$  & Dec=$-$53$^{o}$39$^{'}$26$^{"}$  &       &              \\
2003 July 06  & 2$\times$500s$+$529s$+$3$\times$600s$+$625s      & {\it HST} & WFC/ACS & F775W & 9984 & J.\,D.\,Rhodes \\
2003 July 15  & 4$\times$500s$+$529s$+$589s$+$2$\times$625s      & {\it HST} & WFC/ACS & F775W & 9984 & J.\,D.\,Rhodes \\

  \hline
    \end{tabular}
    \label{dataset_table}
\end{table*}

\section{Multiple populations and cluster mass}
In a recent study, \citet{dondoglio2022a} conducted a survey of multiple populations among M-dwarfs using F110W and F160W data collected with IR/WFC3 on board the {\it HST}. Analyzing the F110W$-$F160W color distribution of M-dwarfs, Dondoglio and collaborators concluded that the properties of multiple populations vary significantly between clusters. Notably, the MS width, $W_{\rm F110W,F160W}$, measured 2.0 magnitudes below the MS knee, exhibits strong correlations with both cluster mass and internal oxygen variation.

\begin{figure} 
  \centering 
 \includegraphics[height=9cm,trim={0.cm 5cm 4.cm 4.5cm},clip]{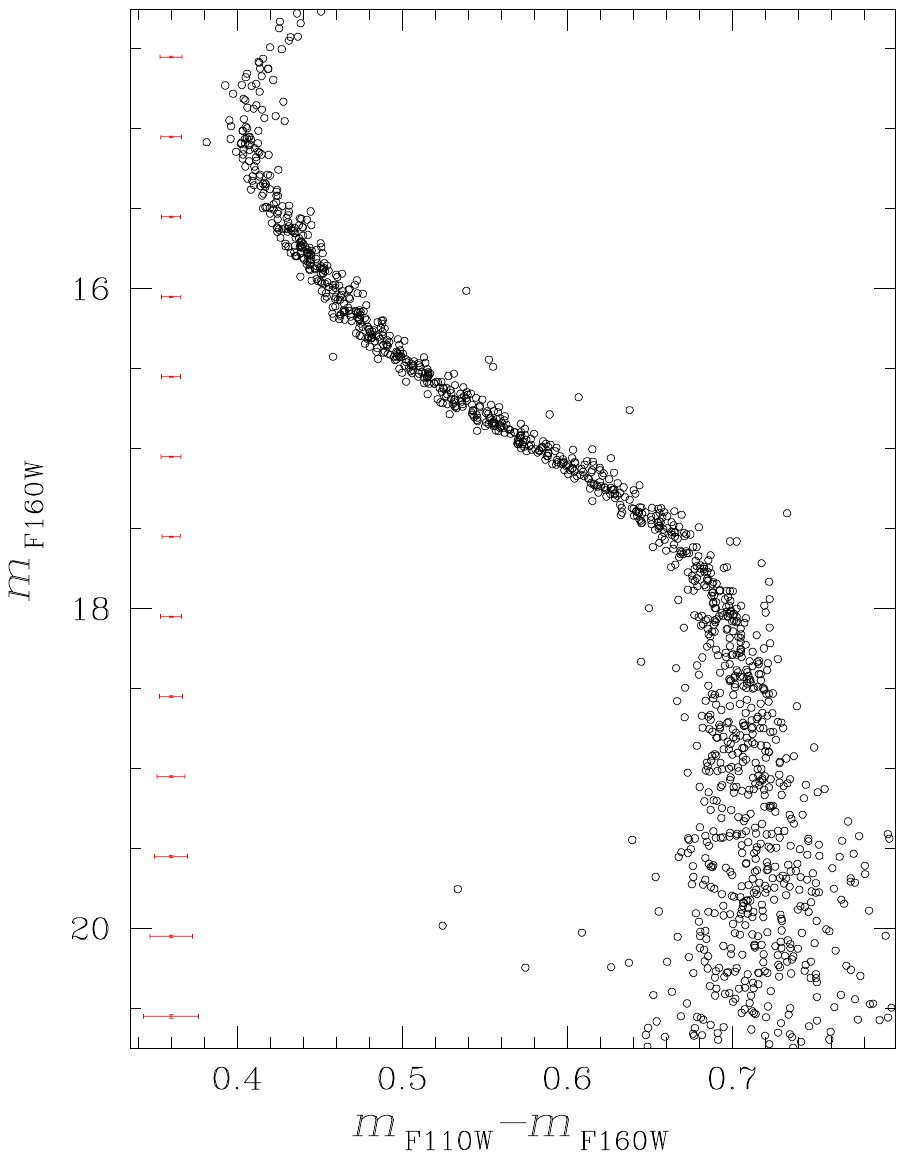}
 \includegraphics[height=4.75cm]{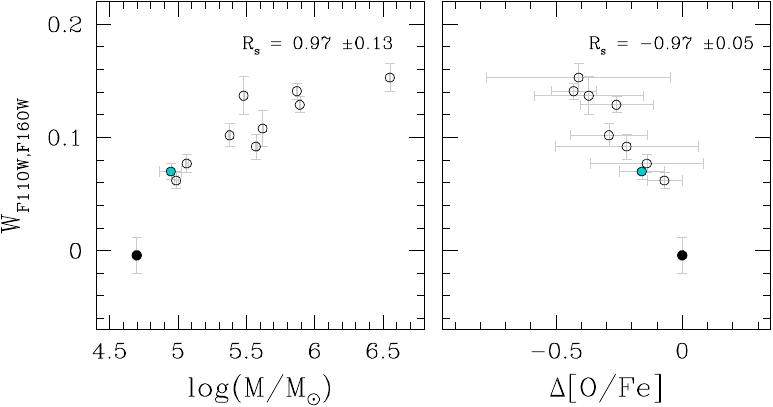}
 \caption{\textit{Top.} $m_{\rm F160W}$ vs.\,$m_{\rm F110W}-m_{\rm F160W}$ CMD for NGC\,6397 stars derived from IR/WFC3 data. \textit{Bottom.} Color width of the M-dwarf sequence, $W_{\rm F110W,F160W}$, as a function of the logarithm of the cluster mass (left) and the internal oxygen variation (right). The black dot represents the open cluster NGC\,6791, NGC\,6397 is shown in azure, and the remaining GCs are displayed with white dots \citep{dondoglio2022a}. The Spearman's rank correlation coefficients are quoted in each panel.}  
  \label{fig:relmass}
\end{figure}

As a byproduct of our research, we utilize the $m_{\rm F160W}$ vs.,$m_{\rm F110W}-m_{\rm F160W}$ CMD derived for central-field stars in NGC\,6397, shown in the upper panel of Figure,\ref{fig:relmass}, to further investigate the relations identified by Dondoglio and collaborators. The results by \citet{dondoglio2022a} are reproduced in the bottom panels of Figure,\ref{fig:relmass}, where white and black dots represent the nine GCs studied by these authors and the open cluster NGC\,6791, respectively.

The MS width of NGC\,6397, determined from the CMD in Figure,\ref{fig:relmass} using the method outlined by \citet{dondoglio2022a}, is represented by an azure dot. The fact that our findings for NGC\,6397 align with the established relations between cluster mass and oxygen variations supports the conclusion of Dondoglio and collaborators that the extent of internal oxygen variation in GCs depends on cluster mass. Similar trends involving internal nitrogen and helium variations among RGB stars have also been reported in previous studies \citep[see][]{milone2017a, milone2018a, lagioia2019a, jang2021a}.
\end{appendix}

\label{lastpage}
\end{document}